\documentclass[a4paper,fleqn,usenatbib]{mnras}

\usepackage{natbib} 
\usepackage{amsmath,amssymb}
\usepackage{epsfig,graphicx}
\usepackage{epstopdf}
\usepackage{subcaption}
\usepackage{bm}
\usepackage{float}
\usepackage[usenames,dvipsnames]{xcolor}
\usepackage{lastpage}

\DeclareGraphicsExtensions{.pdf,.png,.jpg,.eps}
\newcommand{\wgg}{\ensuremath{w_{gg}}}

\newcommand{\mpch}{\ensuremath{h^{-1}\text{Mpc}}}

\def\reff@jnl#1{{\rm#1\/}}

\def\aj{\reff@jnl{AJ}}                  
\def\araa{\reff@jnl{ARA\&A}}            
\def\apj{\reff@jnl{ApJ}}                        
\def\apjl{\reff@jnl{ApJ}}               
\def\apjs{\reff@jnl{ApJS}}              
\def\apss{\reff@jnl{Ap\&SS}}            
\def\aap{\reff@jnl{A\&A}}               
\def\aapr{\reff@jnl{A\&A~Rev.}}         
\def\aaps{\reff@jnl{A\&AS}}             
\def\baas{\reff@jnl{BAAS}}              
\def\jrasc{\reff@jnl{JRASC}}            
\def\memras{\reff@jnl{MmRAS}}           
\def\mnras{\reff@jnl{MNRAS}}            
\def\physrep{\reff@jnl{Phys.Rep.}}
\def\pra{\reff@jnl{Phys.Rev.A}}         
\def\prb{\reff@jnl{Phys.Rev.B}}         
\def\prc{\reff@jnl{Phys.Rev.C}}         
\def\prd{\reff@jnl{Phys.Rev.D}}         
\def\prl{\reff@jnl{Phys.Rev.Lett}}      
\def\pasp{\reff@jnl{PASP}}              
\def\pasj{\reff@jnl{PASJ}}              
\def\skytel{\reff@jnl{S\&T}}            
\def\solphys{\reff@jnl{Solar~Phys.}}    
\def\sovast{\reff@jnl{Soviet~Ast.}}     
\def\ssr{\reff@jnl{Space~Sci.Rev.}}     
\def\nat{\reff@jnl{Nature}}             

\newcommand{\beq}{\begin{equation}}
\newcommand{\eeq}{\end{equation}}
\newcommand{\beqa}{\begin{eqnarray}}
\newcommand{\eeqa}{\end{eqnarray}}

\graphicspath{{figures/}}

\newcommand{\referee}[1]{{{#1}}}
\newcommand{\refereeTwo}[1]{{ {#1}}}

\newcommand{\lcdm}{{$\Lambda$CDM}}
\newcommand{\rcc}{\ensuremath{r_{cc}}}

\newcommand{\DS}{\ensuremath{\Delta\Sigma}}
\newcommand{\SO}{\ensuremath{S_8}} 
\newcommand{\ugm}{\ensuremath{\Upsilon_\text{gm}}}

\title[BOSS$\times$ lensing cosmology]{Cosmological constraints from galaxy-lensing cross correlations using BOSS galaxies with SDSS and CMB lensing}

\author[]{
Sukhdeep Singh$^{1,2}$\thanks{\tt sukhdeep1@berkeley.edu},
Rachel Mandelbaum$^{2}$,
Uro\v{s} Seljak$^1$,
Sergio Rodr\'iguez-Torres$^3$,
\newauthor
An\v{z}e Slosar$^4$\\
$^1$Berkeley Center for Cosmological Physics, University of California, Berkeley, CA 94720, USA\\
$^2$McWilliams Center for Cosmology, Department of Physics, Carnegie Mellon University, Pittsburgh, PA 15213, USA\\
$^3$Departamento de F\'isica Te\'orica M8, Universidad Aut\'onoma de Madrid (UAM), Cantoblanco, E-28049, Madrid, Spain\\
$^4$Physics Department, Brookhaven National Laboratory, Upton, NY 11973, USA}

\date{\today}
\date{Accepted XXX. Received YYY; in original form ZZZ}
\pubyear{2018}

\begin{document}
\label{firstpage}
\pagerange{\pageref{firstpage}--\pageref{LastPage}}
\maketitle

\begin{abstract}
	We present cosmological parameter constraints based on a joint modeling of galaxy-lensing cross correlations and galaxy
	clustering measurements in the SDSS, marginalizing over
	small-scale
	modeling uncertainties using mock galaxy catalogs, without explicit modeling of galaxy bias. We show that our modeling
    method is robust
	to the impact of different choices for how galaxies occupy dark matter halos and to the impact of baryonic physics (at the $\sim2\%$ level
	in cosmological parameters)
    and test for the impact of covariance on the likelihood analysis and of the survey window function on the theory
	computations.
	Applying our results to the measurements using galaxy samples from BOSS and lensing measurements using shear from SDSS galaxies and CMB
	lensing from Planck, with conservative scale cuts, we obtain
        $\SO\equiv\left(\frac{\sigma_8}{0.8228}\right)^{0.8}\left(\frac{\Omega_m}{0.307}\right)^{0.6}=0.85\pm0.05$ (stat.)
        using LOWZ $\times$ SDSS galaxy
    lensing, and  $\SO=0.91\pm0.1$ (stat.) using combination of LOWZ and CMASS $\times$
	Planck CMB lensing.
    We estimate the
    systematic uncertainty in the galaxy-galaxy lensing measurements to be $\sim6\%$ (dominated by photometric redshift uncertainties) and in
    the galaxy-CMB lensing
    measurements to be $\sim3\%$, from small scale modeling uncertainties including baryonic physics. 
\end{abstract}


\section{Introduction}\label{sec:intro}
The \lcdm\ cosmological model has been successful in explaining a wide array of cosmological observables. However, the model
includes as yet poorly understood cold dark matter (CDM) and the cosmological constant, $\Lambda$ (or dark energy in general). Both of these components
affect the expansion history and the growth of structure in the Universe, which provides a promising avenue to probe them.

Gravitational lensing, the perturbations to the path of photons by the gravitational effects of matter, has emerged as an important probe of the
geometry and the growth of structure in the universe  \citep{Bartelmann2001,Weinberg2013,Kilbinger2015,Mandelbaum2017}.
In the weak regime, lensing effects of the (foreground) matter introduce small and coherent
perturbations in
the light profiles
of background objects. By measuring the correlations between shapes or sizes of background galaxies (or CMB perturbations when using
CMB lensing), we can infer the structure in the distribution of foreground matter. By selecting sources at different distances or redshifts, weak lensing
also enables the study of growth and evolution of structure in the universe. Several studies have used the correlations in the shapes of galaxies (cosmic
shear, or shear-shear correlations) to probe the growth history of the Universe \citep[e.g., ][]{Semboloni2006,Massey2007,Kitching2014,Hildebrandt2016,Uitert2017,DES2017comb}
and more recently there have been similar studies using the lensing of
CMB \citep[e.g., ][]{Das2011,Planck2013lensing,Planck2015lensing,Omori2017,Sherwin2017}.

The cross correlations between the galaxies and the lensing maps provide another avenue to probe the distribution of matter around galaxies. The
galaxy-galaxy lensing cross correlation have been used by several studies to probe the connection between the galaxies and their host halos
\citep[e.g.,][]{Hoekstra2004,Mandelbaum2006,Heymans2006,Tinker2012,Leauthaud2012,Uitert2012,Gillis2013,Velander2014,Sifon2015,Hudson2015,Uitert2016,Dvornik2018}.
Combined with galaxy clustering, these measurements also provide a measure of the matter correlations (matter power spectrum), independent of
the galaxy bias. These measurements have been widely used to measure the cosmological parameters, especially the amplitude of the matter power spectrum
quantified by $\sigma_8$ at various redshifts \citep{Seljak2005,Baldauf2010,Mandelbaum2013,More2015,Kwan2016,Buddendiek2016}.
Similar studies have also been performed using CMB lensing \citep{Hand2015,Giannantonio2016,Kirk2016,Singh2017cmb}.
Furthermore, combined with other probes of large scale structure, such as redshift space distortions, these measurements can also be used to test
theories of gravity \citep[e.g.,][]{Zhang2007,Reyes2010,Blake2016,Alam2016,Pullen2016,Singh2018eg}.

When extracting cosmological information from galaxy clustering and galaxy-galaxy lensing, an important limiting factor is the difficulty in
modeling the correlations at smaller spatial scales. Models for correlations in the matter density field that are based on perturbation theory tend to
fail at small scales
 \citep[see][for review]{Bernardeau2002}
 and it is also difficult to predict the non-linear galaxy bias and the stochastic
cross correlations between the galaxies and matter. A
conservative solution is to only use the measurement on large scales where the model is reliable,
though this can involve a considerable loss of information from small scales, where having more modes results in better signal-to-noise measurements.
To model the small scales, one must resort to
modeling based on N-body simulations using tools such as Halo Occupation and Distribution (HOD)
models, which provide a framework to populate halos with galaxies \citep[see][for review]{Cooray2001}.
HOD-based modeling has been
used by \cite{Seljak2005,Cacciato2009,Cacciato2013,More2015}
 to derive cosmological parameter constraints \citep[see also][]{Wibking2017}.

While HOD modeling provides an effective and potentially flexible way of
modeling the observables, it requires assumptions about the relationships between the galaxies and their host halos and then marginalizes over model
parameters describing those relationships.
These assumptions in turn can affect the cosmological inferences.  While a more flexible HOD can in principle provide sufficient model freedom to reduce
 biases substantially, having a very flexible HOD can result in loss of information.  In contrast, having too inflexible
model can result in overly optimistic and potentially biased inferences.

Another approach to extracting information from smaller scales is to model the matter power spectrum to smaller scales using emulators trained on N-body
simulations but also remove the information from small scales that are strongly affected by the stochastic relation between galaxies and matter.
One such approach, suggested by \cite{Baldauf2010}, involves removing the small scale information from the lensing observables and then modeling the
remaining linear and quasi-linear scales using a simulation-based emulator or higher order perturbation theory. This approach was successfully applied
by \cite{Mandelbaum2013,Singh2017cmb} to derive cosmological parameters and by several papers to test gravity using the $E_G$
parameter \citep{Reyes2010,Blake2016,Alam2016,Singh2018eg}.

In this work, we develop the methodology to derive the cosmological parameters from the combination of galaxy clustering and galaxy-lensing cross
correlations following the approach of \cite{Baldauf2010}, but without having to explicitly model the galaxy bias. We model the galaxy-matter cross correlation
coefficient (which can be thought of as the
difference in non-linear bias between galaxy-galaxy and galaxy-matter correlations) using simulations.
We will also extend the results using the estimator proposed by \cite{Baldauf2010}
down to smaller scales than has been done before.

We begin by describing the lensing and clustering formalism as well as the estimators used to measure the
signals in section~\ref{sec:formalism} followed by the description of modeling schemes in
section~\ref{sec:modeling}. We describe the datasets and mock datasets used for measurements in
section~\ref{sec:data}. Results from fitting the measurements from both data and mock galaxy catalogs are presented
in section~\ref{sec:results} and we conclude in section~\ref{sec:conclusions}.

Throughout we use $\Omega_m=0.307,\sigma_8=0.8228,h=0.6777,n_s=0.96$ \citep{Planck2014} 
as our fiducial cosmology, and express comoving distances in units of \mpch.

\section{Formalism and Methodology}
\label{sec:formalism}
	In this section we review the basic formalism of galaxy clustering and galaxy-lensing cross correlation measurements as well as the
	estimator used for the measurements.
	\subsection{Weak Lensing}
		The gravitational lensing effects of the foreground matter distribution introduce small but coherent distortions in
		the light profiles
 		of the background objects \citep[see][for review]{Bartelmann2001,Weinberg2013}.
		The gravitational lensing measurements are sensitive to the lensing potential defined as
		\begin{equation}
			\Phi_L=\int d\chi_l\frac{f_\kappa(\chi_s-\chi_l)}{f_\kappa(\chi_s)f_\kappa(\chi_l)}\Psi(f_
			\kappa(\chi_l)\vec\theta,\chi_l),
		\end{equation}
		where the  Weyl potential $\Psi=\psi+\phi$, and $\phi$ and $\psi$ are the Newtonian and curvature potentials. $\chi_s$
        and $\chi_l$ are line-of-sight comoving distances to source and lens redshifts and $f_k(\chi)$ are the transverse
        comoving distances.
        Within
		GR,
         $\phi=\psi$ and $\nabla^2\phi=4\pi G\rho_m$.
		The Jacobian of relating image to source locations is
    \begin{align}
      A_{ij}&=\frac{\partial(\theta_o^i-\delta\theta_o^i)}{\partial\theta_o^j},\\
      A_{ij}&=\delta_{ij}-\frac{\partial^2\Phi_l}{\partial\theta_o^i\partial\theta_o^j},
    \end{align}
    \[
    	A=
	      \begin{bmatrix}
    	    1-\kappa-\gamma_1 & -\gamma_2\\
        	-\gamma_2 & 1-\kappa+\gamma_1
	      \end{bmatrix}
    \].
		The primary weak gravitational lensing
		observables are the convergence $\kappa$ and the shear $\gamma=\gamma_1+\mathrm{i}\gamma_2$, defined as
		\begin{align}
	      \kappa&=-\frac{1}{2}\nabla_\perp^2\Phi_L\leftrightarrow\frac{1}{2}k_\perp^2\Phi_L,\\
    	  \gamma_1&=\frac{1}{2}(\nabla_{1,1}^2-\nabla_{2,2}^2)\Phi_L\leftrightarrow\frac{1}{2}(k_{\perp,1}^2-k_{\perp,2}^2)\Phi_L,\\
	      \gamma_2&=\frac{1}{2}(\nabla_{1,2}^2)\Phi_L\leftrightarrow\frac{1}{2}(k_{\perp,1}k_{\perp,2})\Phi_L,
    	\end{align}
		where derivatives are with respect to the projected (perpendicular to line of sight) coordinates and \referee{we have used $\leftrightarrow$ to denote equivalent Fourier space counterparts.}

	\subsubsection{Galaxy lensing cross correlations}

Galaxy lensing cross correlations measure the projected surface mass density, $\Sigma$, around the
		galaxies, which is related to convergence and shear as
		\begin{align}
			&\kappa(r_p)=\frac{\Sigma(r_p)}{\Sigma_{c}},\\
			&\gamma_t(r_p)=\frac{\overline{\Sigma}(<r_p)-\Sigma(r_p)}{\Sigma_{c}}.
		\end{align}
		$\overline{\Sigma}(<r_p)$ is the mean surface mass density within the radius $r_p$ with the critical density, $\Sigma_c$
		defined as
		\begin{equation}\label{eqn:sigma_crit}
    		\Sigma_c=\frac{c^2}{4\pi G}\frac{f_k(\chi_s)}{(1+z_l)f_k(\chi_l) f_k(\chi_s-\chi_l)}.
	    \end{equation}

		$\Sigma$ is related to the projected galaxy-matter cross correlations as
		\begin{equation}
    		\Sigma(r_p)=\bar\rho_m\int \mathrm{d}\Pi \, \xi_{gm}(r_p,\Pi)=\bar\rho_m w_{gm}(r_p).
	    \end{equation}
		The matter-galaxy correlation function, $\xi_{gm}$, can further be related to the matter power spectrum as
		\begin{align}
			\xi_{gm}(r_p,\Pi)=b_g(r_p) r_{cc}(r_p)&\int \mathrm{d}z W_L(z)\int
            \frac{\mathrm{d}^2k_\perp dk_z}{(2\pi)^3}\nonumber\\&\times P_{\delta
			\delta}(\vec{k},z)e^{\mathrm{i}(\vec{r}_p.\vec{k}_\perp+\Pi k_z)},
            \label{eqn:xi}
	    \end{align}
		 where $b_g(r_p)$ is the galaxy bias and $\rcc(r_p)$ is the galaxy-matter cross correlation coefficient \referee{and are defined as}
		 \begin{align}
		 	b_g(r_p)&=\sqrt{\frac{\xi_{gg}(r_p)}{\xi_{mm}(r_p)}},\\
			\rcc(r_p)&=\frac{\xi_{gm}(r_p)}{\sqrt{\xi_{gg}(r_p)\xi_{mm}(r_p)}}.
		 \end{align}
		 The weight function \referee{$W_L(z)$ denotes the effective weight/contribution from lens galaxies at 
		 each redshift to 
		 the measured signal} and \refereeTwo{is given as
		 \begin{equation}
		 	W_L(z_l)=p(z_l)\int^\infty_{z_l} dz_s p(z_s) \Sigma_c(z_l,z_s) w_{ls}
		\end{equation}
		 where $p(z_l)$ and $p(z_s)$ are the redshift distributions of lens and source galaxies (in the case of galaxy lensing) respectively, 
		 $\Sigma_c$ is the critical density (lensing kernel) for the lens-source pair and
		 $w_{ls}$  is the lens-source pair weights 
		 used in the estimators
		 when measuring the signal (see Section~\ref{ssec:estimators}).
		 }
		 
	\subsection{Galaxy Clustering}
		The two-point
		correlation function of galaxies is given by
		\begin{align}
			\xi_{gg}(r_p,\Pi)=&b_g^2(r_p)\int \mathrm{d}z W(z)\int \frac{\mathrm{d}^2k_\perp\mathrm{d}k_z}{(2\pi)^3}\nonumber\\
			&\times P_{\delta\delta}(\vec{k},z)(1+\beta\mu_k^2)^2 e^{\mathrm{i}(\vec{r}_p.\vec{k}_\perp+\Pi k_z)}\label{eqn:xi}.
	    \end{align}
		The Kaiser factor, $(1+\beta\mu_k^2)$ accounts for the redshift space
		distortions \citep{Kaiser1987}, where $\beta=f(z)/b_g$, $f(z)$ is the linear growth rate factor at redshift $z$ 
		and
		$\mu_k=k_z/k$. The weight function\footnote{\refereeTwo{Note that here we assume that the correlations for a given galaxy only contain contributions 
		from modes at the same redshift as the galaxy, which then allows us to separate the integrals over redshift and $k_z$}.} \referee{$W(z)$
		denotes the effective weight/contribution of galaxies at each 
		redshift to 
		 the measured signal and} is given by \citet{Mandelbaum2011}
			as
			\begin{equation}
				W(z)=\frac{p(z)^2}{\chi^2 (z)\mathrm{d}\chi/\mathrm{d}z} \left[\int \frac{p(z)^2}{\chi^2
				(z)\mathrm{d}\chi/\mathrm{d}z} \mathrm{d}z\right]^{-1}.
			\end{equation}
		$p(z)$  is the redshift probability distribution for the galaxy sample.

		Finally we integrate $\xi_{gg}$ over the line-of-sight separation to get the projected correlation
		function
		\begin{equation}
			w_{gg}(r_p)=\int\limits_{-\Pi_\text{max}}^{\Pi_\text{max}}\mathrm{d}\Pi \, \xi_{gg}(r_p,\Pi).
		\end{equation}
		The choice of $\Pi_\text{max}$ depends on the impact of redshift space distortions (RSD), which leads to a need for large $\Pi_\text{max}$ especially
		when going to large $r_p$, and the
        increase of
		noise in the final measurement, which increases with $\Pi_\text{max}$. We choose
        $\Pi_\text{max}=100\mpch$ and then apply the corrections
        for
		RSD to the measured projected correlation function using the Kaiser factor.
		As was shown in \cite{Singh2018eg} this correction is much smaller than the statistical
        errors on the scales we use and has a negligible effect on our final results.

	\subsection{Estimators}\label{ssec:estimators}
		In this section we present the estimators used for measuring various signals. For all of the measurements, we use
		100 approximately equal-area ($\sim8$ degrees on a side) jackknife regions to obtain the jackknife mean and
		covariance \citep[see][for more details]{Singh2017cmb,Singh2017cov}.

		\subsubsection{Galaxy-Galaxy Lensing}
			For galaxy-galaxy lensing, we measure the excess surface density $\Delta \Sigma$ as
    		\begin{equation}
    	    	\widehat{\Delta \Sigma}(r_p)=B_L(1+m_{\gamma})\left[\frac{\sum_{ls}w_{ls}e_t^{(ls)}\Sigma_\text{crit}^{(ls)}}{2\mathcal R\sum_{rs}w_{rs}}-
				\frac{\sum_{rs}w_{rs}e_t^{(rs)}\Sigma_\text{crit}^{(rs)}}{2\mathcal R\sum_{rs}w_{rs}}\right].
		       	\label{eq:delta_sigma_estimator}
    		\end{equation}

            The summation is over all lens-source (ls) and random-source (rs) pairs. $e_t$ is the tangential ellipticity measured in the
            lens-source frame, $m_\gamma$ is the
            multiplicative bias in our shear estimates and
            $\mathcal R$ is the responsivity factor to convert the ensemble average of ellipticities to shear.
            $B_L$ is the calibration factor to account for bias due to photometric redshift of source galaxies.
            The weight $w_{rs}$ in the denominator of Eq.~\eqref{eq:delta_sigma_estimator}
            accounts for the source galaxies associated with the lens, which do not contribute any shear but are counted in
            the total weights. The correction factor for this effect $w_{ls}/w_{rs}$ is usually called the boost
            factor \citep{Sheldon2004,Mandelbaum2005}. Finally, we also subtract the shear signal measured around the
            random points for two reasons: to remove systematics that can give some spurious shear signal at large scales, and for the optimal covariance
            of the final measurements \citep{Singh2017cov}.
            The weight $w_{ls}$ for each lens-source pair is defined as
            \begin{equation}\label{eq:wls}
               w_{ls}=\frac{w_l\Sigma_c^{-2}}{\sigma_\gamma^2+\sigma_{SN}^2}.
            \end{equation}
            The $\Sigma_c^{-2}$ in the weight is required for the optimal estimator in the shape
            noise-dominated regime \citep{Sheldon2004}, $w_l$ is the weight assigned to lens
            galaxies (systematic weights for BOSS galaxies, see
			section~\ref{ssec:data_Boss}), $\sigma_\gamma$ is the measurement noise for the galaxy
            shape and $\sigma_{SN}$ is the shape noise.

        \subsubsection{Galaxy-CMB lensing}
        From CMB lensing maps, we get estimates of the convergence, from which we measure the projected surface mass density around galaxies as
         \citep{Singh2017cmb}
		    \begin{equation}
    	        \widehat{\Sigma}(r_p)=\frac{\sum_{lp}w_{lp}\kappa_{p}\Sigma_{c,{*}}}{\sum_{lp}w_{lp}}
				-\frac{\sum_{Rp}w_{Rp}\kappa_{p}\Sigma_{c,{*}}}{\sum_{Rp}w_{Rp}},
        	   \label{eq:sigma_cmb}
    		\end{equation}
			where the summation is over all the lens-pixel (pixels of CMB convergence map) pairs at separations
			$r_p\in[r_{p,min},r_{p,max}]$ at the lens redshift and the signal around random points is subtracted to achieve a more optimal measurement
			 \citep{Singh2017cov}. $\Sigma_{c,*}$ is $\Sigma_\text{crit}$ with CMB as the source.

            Similar to the galaxy-galaxy lensing case, the weight for each lens-pixel pair is given by
    		\begin{equation}
            	w_{lp}=w_l\Sigma_{c,*}^{-2}.
        	\end{equation}

		\subsubsection{Galaxy Clustering}
			We use the Landy-Szalay \citep{landy1993} estimator to compute the galaxy two-point correlation function:
            \begin{equation}
               \xi_{gg}(r_p,\Pi)=\frac{DD-2DR+RR}{RR},
            \end{equation}
            where $D$ is the weighted galaxy sample, with weights defined in Section~\ref{ssec:data_Boss}
			and $R$ is for the random point sample (corresponding to the weighted galaxy catalog).
            The projected correlation function is obtained by integrating over the bins in line-of-sight separation, 
            $\Pi$
            \begin{equation}
               w_{gg}(r_p)=\sum_{-\Pi_\text{max}}^{\Pi_\text{max}}\Delta\Pi\, \xi_{gg}(r_p,\Pi).
            \end{equation}
            We use $\Pi_\text{max}=100\mpch$, with 20  bins of size $\Delta\Pi=10\mpch$. \referee{$\Pi_\text{max}=100\mpch$ is large enough so 
            that the effects of redshift space distortions are small and avoids including very large line of sight scales which contribute little 
            to the signal but increase the noise \citep[see more discussion in ][]{Singh2018eg} }

	\subsection{Removing small scale information}
    	\label{ssec:estimator_upsilon}

		We use the estimator suggested by \cite{Baldauf2010} to remove the difficult-to-model small-scale information from \DS\
		 \begin{equation}
		 	\Upsilon_{gm}(r_p,r_0)=\Delta\Sigma(r_p)-\left(\frac{r_0}{r_p}\right)^2\Delta\Sigma(r_0).
		 \end{equation}
		 $\Sigma(r_p)$ and \wgg\ ($\Sigma_{gg}=\overline{\rho}_m\wgg$)
         can also be converted to $\Upsilon$ using the relations between
		 $\Sigma$ and $\Delta\Sigma$ and corresponding theoretical calculations can be done. In addition to removing
		 the small scale information, \cite{Baldauf2010} also showed that $\Upsilon$ reduces the impact of sample variance
          and the redshift space distortions in the projected correlation functions. We will use
		 $\Upsilon$ as well as \DS\ with  a model inspired from $\Upsilon$ in our analysis. To compute $\Upsilon$, we also need to choose the
         scale $r_0$ below which all the information is removed. We will try few different values of $r_0$, with the aim of choosing to smallest
         possible $r_0$ values for which we can obtain unbiased results on the cosmological parameters in the mock galaxy catalogs.

\section{Modelling}\label{sec:model}
\label{sec:modeling}
	In this section we describe three different methods to model the observables, $\ugm$ or $\DS_{gm}$, while removing 
	the information from small
	scales. In short, we will model the galaxy-lensing cross correlations in terms of galaxy and matter clustering.
	where 
	matter clustering is
    \referee{predicted using analytical model} 
    and galaxy clustering is directly taken from the measurements. 
    We will
    use the non-linear matter power spectrum computed using halofit \citep{Takahashi2012} using {\sc CLASS} \citep{Lesgourgues2011}.
    In the
    current approach, we will be ignoring the covariance of galaxy clustering since the galaxy correlation functions are measured to much better
    precision than the galaxy-lensing cross correlations.
    We will later test these models on mock galaxy catalogs, which will also allow us to choose the values of the cutoff scale $r_0$.

	\subsection{Modelling $\Upsilon_{gm}$}
    \label{ssec:modeling_upsilon}

		We write the model for galaxy-matter correlations $\Upsilon_{gm}$ in terms of the clustering of galaxies, $\Upsilon_{gg}
		$ and matter, $\Upsilon_{mm}$, as:
		\begin{equation}
			\Upsilon_{gm}=\bar{\rho}_mr_{cc}^{(\Upsilon)}\sqrt{\Upsilon_{mm}\Upsilon_{gg}},
			\label{eq:upsgm_predict}
		\end{equation}
		where $r_{cc}^{(\Upsilon)}$ is the cross correlation coefficient between the galaxies and matter when using the $\Upsilon$ estimator.
		We also assume $\Upsilon_{gg}$
		and $\Upsilon_{mm}$ are in units of length (no factors of density) and hence multiply with $\bar{\rho}_m$ to get the
		units of projected surface mass density.

\begin{figure}
			\centering
			\includegraphics[width=\columnwidth]{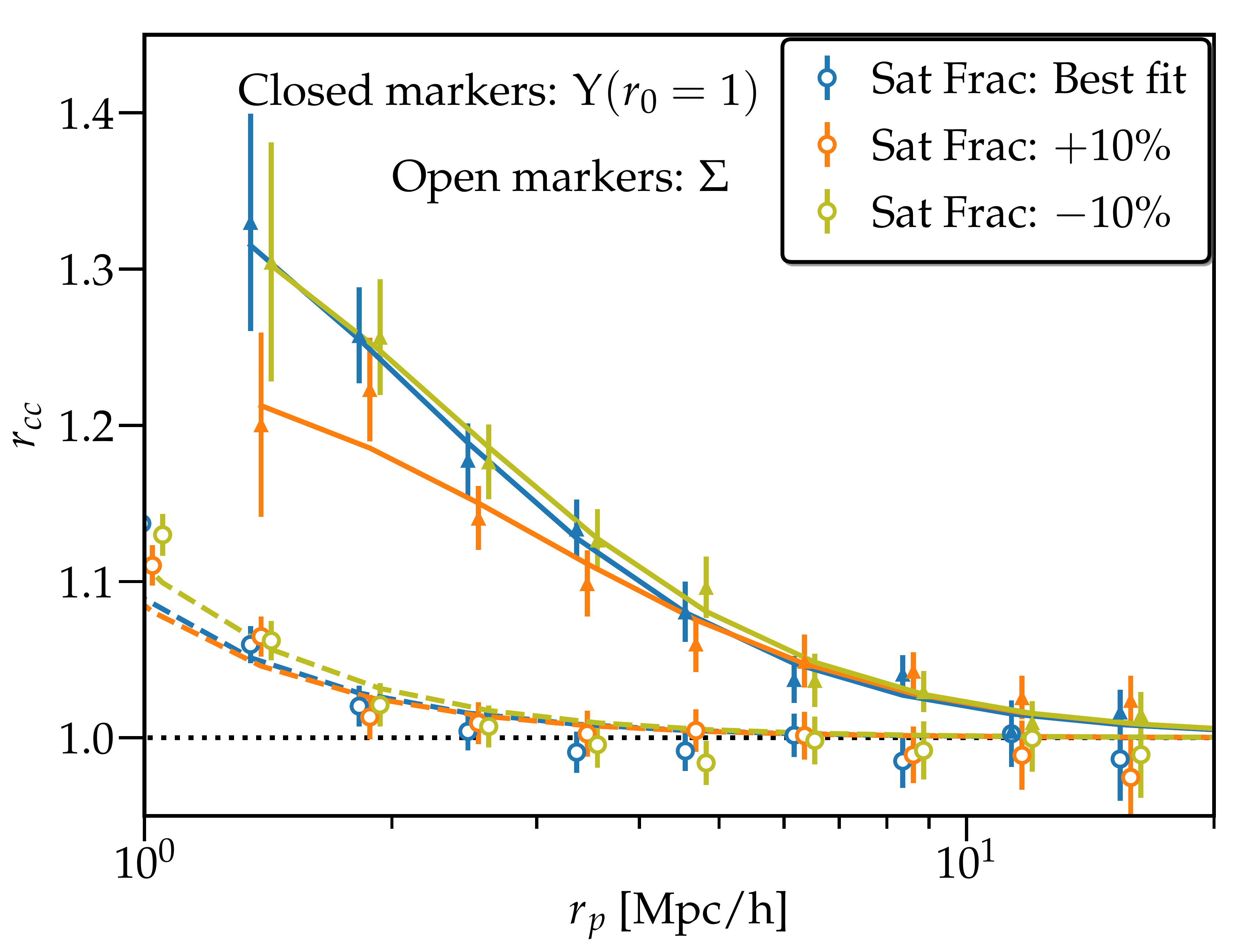}
                        \caption{Galaxy-matter cross-correlation coefficient \rcc\ estimates from mock galaxy catalogs with
                          different satellite fractions (see
                          Section~\ref{ssec:z_mocks}) and parametric model
                          fits from eq.~\eqref{eq:rcc_predict}. Closed triangles show the \rcc\ estimates using the $\Upsilon$
                          estimator and open circles are using the $\Sigma$.
                          This plot illustrates that the
                          functional form adopted is sufficiently flexible to describe the
                          behavior found in simulations, but we stress
                          that the detailed form of $\rcc$ is  not important, as we
                          find small shifts ($\lesssim 0.3\sigma$)
                          in our results for cosmological
                          parameters even when
                          assuming $\rcc=1$ on all scales used.
}
\label{fig:rcc_fits}
\end{figure}

		We model $r_{cc}$ as
		\begin{equation}
			r_{cc}(r_p)=1+\frac{a}{r_p^2+b},
			\label{eq:rcc_predict}
		\end{equation}
		where $a,b$ are free parameters to be fit during the likelihood analysis to obtain cosmological parameters.
        		We also
                estimate the range of values $r_{cc}$ can take as
                a function of scale in mock galaxy catalogs and put (uniform)
                priors to constrain $r_{cc}$ to stay within that range
                (note that the prior is on $r_{cc}$ as a function of scale
                and not on $a,b$, see Fig.~\ref{fig:rcc_comp}). We
                note that the choice of functional form in
                Eq.~\eqref{eq:rcc_predict} is arbitrary,
                chosen to fit the shape of \rcc\ observed in
                simulations.  In Fig.~\ref{fig:rcc_fits} we show that this particular functional form is sufficiently
				flexible to fit \rcc\ estimates from the variety of mock galaxy catalogs described in Section~\ref{ssec:z_mocks}.
								We have also tested our results with some other
                functional forms (not shown in this paper) with similar shapes, and our results are not
                affected by this particular choice.

	\subsection{Modelling $\Delta\Sigma$}
    \label{ssec:modeling_DS}

		\subsubsection{Model using $\ugm$ ($\Delta\Sigma_{gm}\leftarrow\Upsilon_{gm}$)}
			Using the prediction of \ugm\ from Eq.~\eqref{eq:upsgm_predict}, we can predict $\DS_{gm}$ as
			\begin{equation}
				\Delta\Sigma_{gm}(r_p)=\Upsilon_{gm}(r_p)+\frac{r_0^2}{r_p^2}\Delta\Sigma_0,
			\end{equation}
			where $\DS_0$ is an additional free parameter to be marginalized over.

		\subsubsection{Model using $\Sigma$ $(\Delta\Sigma_{gm}\leftarrow\Sigma_{gm})$}
			One of the motivations to use $\Upsilon$ parameter is to remove the information from scales where the $r_{cc}$
            value
			deviates significantly from $1$. To include some scales where $r_{cc}$ deviates from 1, one can also
            predict
			\DS\ directly from $\Sigma$, where $r_{cc}$ values are more localized in $r_p$ \referee{(i.e. deviate from 1 at smaller $r_p$)}
            compared to \DS\ and $\Upsilon$.
			In this case we
			begin by predicting $\Sigma_{gm}$, 
			\begin{equation}
				\Sigma_{gm}=r_{cc}^{(\Sigma)}\sqrt{\Sigma_{mm}\Sigma_{gg}},
			\end{equation}
			where we use same model as in Eq.~\eqref{eq:rcc_predict} to predict $r_{cc}^{(\Sigma)}$ and also put priors on $r_{cc}^{(\Sigma)}$ based
			on the range of values observed in mock galaxy catalogs.

			$\Sigma_{gm}$ can then be converted to \DS\ as
			\begin{equation}
				\Delta\Sigma_{gm}(r_p)=\frac{1}{r_p^2}\int_{r_0}^{r_p}\mathrm{d}r_p'r_p'\Sigma_{gm}+
									\frac{1}{r_p^2}\Sigma_{0}-\Sigma_{gm}(r_p).
			\end{equation}
			Notice that we have restricted the integrals over predicted $\Sigma_{gm}$ on scales $>r_0$ and also introduced the
			parameter $\Sigma_{0}$, which is the integrated $\Sigma_{gm}$ over scales $<r_0$.
			\begin{equation}
				\Sigma_{0}=\int_{0}^{r_0}dr_p'r_p'\Sigma_{gm}=r_0^2(\DS_{gm}(r_0)+\Sigma_{gm}(r_0)).
			\label{eq:sigma0_DS0}
			\end{equation}

\section{Data}\label{sec:data}
	\subsection{SDSS}
		 The SDSS survey \citep{1998AJ....116.3040G,2000AJ....120.1579Y,2001AJ....122.2129H, 2004AN....325..583I,1996AJ....111.1748F,
         2002AJ....123.2121S,2001AJ....122.2267E,Gunn2006,
		  2002AJ....123.2945R,2002AJ....124.1810S} was an
                imaging and spectroscopic survey, which has imaged approximately quarter
		of the sky, and has produced imaging catalogs
                \citep{Lupton2001,
		  2003AJ....125.1559P,2006AN....327..821T,2009ApJS..182..543A,2011ApJS..193...29A,2008ApJ...674.1217P}
                which have also been used for target selection for
                the spectroscopic follow up in the BOSS survey as well as generation of the shear catalogs described in this section.

		\subsubsection{SDSS-III BOSS}
		\label{ssec:data_Boss}

		As tracers for the galaxy density field, we use SDSS-III BOSS
        \citep{Blanton:2003,Bolton:2012,Ahn:2012,Dawson:2013,Smee:2013} data
        release 12
            \citep[DR12;][]{Alam2015}
			LOWZ galaxies, in the redshift range $0.16<z<0.36$ and CMASS galaxies in the redshift
            range  $0.43<z<0.7$. We apply the systematic weights for the galaxies when computing
            both galaxy clustering and the galaxy lensing cross correlations ($w_l$ as defined in section~\ref{ssec:estimators}), where the weights are
            given by \citep{Ross2012}
            \begin{equation}
            	w_l=w_\text{sys}(w_{no-z}+w_{cp}-1),
            \end{equation}
            where $w_\text{sys}$ weights correct for the variations in the selection
            function on the
            sky
            (important for CMASS) and $w_{no-z}$, $w_{cp}$ correct for missing redshifts due to
            failure to obtain redshift (no-z) or fiber collisions for close pairs, $cp$. Impact of
            these weights on galaxy-lensing cross correlations was studied in
            \cite{Singh2017cmb}.

		\subsubsection{Re-Gaussianization Shapes}

		The shape sample used to estimate the shear
		is described in more detail in \cite{Reyes2012}. Briefly, these shapes are measured
		using the re-Gaussianization technique developed by \cite{Hirata2003}. The
		algorithm is a modified version of ones that use ``adaptive moments'' (equivalent to fitting
        the light intensity profile to an elliptical Gaussian), determining shapes of the
        PSF-convolved galaxy image based on adaptive moments and then correcting the resulting
        shapes based on adaptive moments of the PSF.   The re-Gaussianization method involves
        additional steps to correct for  non-Gaussianity of both the PSF and the galaxy surface
        brightness profiles \citep{Hirata2003}. The components of the distortion are defined as
		\begin{equation}\label{eqn:distortion}
			(e_+,e_\times)=\frac{1-(b/a)^2}{1+(b/a)^2}(\cos 2\phi,\sin 2\phi),
		\end{equation}
		where $b/a$ is the minor-to-major axis ratio and $\phi$ is the position angle of the major
		axis on the sky with respect to the RA-Dec coordinate system. The ensemble average of the
        distortion is related to the shear as
		\begin{align}
			\gamma_+,\gamma_\times&=\frac{\langle e_+,e_\times\rangle}{2\mathcal
			R};\label{eqn:regauss_shear}\\
			\mathcal R&=1-\frac{1}{2}\langle e_{+,i}^2+e_{\times,i}^2-2\sigma_i^2\rangle\label{eq:R},
		\end{align}
		where $\sigma_i$ is the per-component measurement uncertainty of the galaxy distortion, and
		${\mathcal R\approx0.87}$ is the shear responsivity representing the response of an ensemble of
        galaxies with some intrinsic distribution of distortion values
		to a small shear \citep{Kaiser1995,Bernstein2002}.

	\subsection{Redshift-dependent mock galaxy catalogs}
		\label{ssec:z_mocks}
        To get a prior on \rcc\ models as well as to test our models, we use the mock galaxy catalogs
				described here.

The clustering evolution and the abundance of galaxies as a function of redshift are modeled in light-cone catalogues drawn from the BigMultiDark simulation \citep[BigMDPL;][]{Klypin2014}. This 2.5 $h^{-1}$ Gpc simulation with	3840$^3$ particles  adopts a $\Lambda$CDM model using the Planck 2013 cosmological parameters \citep{Planck2014}. In order to identify dark matter halos, the simulation implements the Robust Overdensity Calculation using K-Space Topologically Adaptive Refinement (\textsc{RockStar}) halo finder \citep{Behroozi2013}. The BigMDPL simulation has a volume large enough to construct a light-cone in the whole redshift range covered by the BOSS sample ($0.16<z<0.7$) with an area of $10,206$ deg$^2$. Additionally, this simulation was designed to resolve halos that host CMASS galaxies \citep[$M_h>2.5\times10^{12}M_\odot/h$,][]{Shan2017}, being 95\% complete at $M_{vir}=2\times 10^{12}M_\odot/h$. It allows us to reproduce the observed clustering without including halos from the incomplete region of this simulation. The mean halo mass for the
LOWZ like sample $\sim3.1\times10^{13}M_\odot/h$ and for CMASS sample is $\sim2.4\times10^{13}M_\odot/h$, which are larger than the masses estimated in \cite{Singh2017cmb} using weak lensing, $M_\text{halo}\sim10^{13}M_\odot/h$ for the LOWZ sample using SDSS galaxy lensing (note that the masses obtained using CMB lensing in \citealt{Singh2017cmb} are not reliable due to the resolution of the Planck CMB lensing map).

		Galaxies are assigned to dark matter halos using a \referee{Sub Halo} Abundance Matching \citep[SHAM;][]{Kravtsov2004,Conroy2006,Reddick2013} technique. This process is implemented by using the SUrvey GenerAtoR code \citep[\textsc{sugar;}][]{Rodrigueztorres2016}. The scatter between dark matter halos and galaxies is fixed to reproduce the \referee{projected correlation function} of the BOSS data at different redshifts. \referee{Additionally, we fix the number density at each redshift in order to replicate the observed radial selection function.} We perform the method using the maximum circular velocity over the whole history of the halo ($V_{peak}$) as the proxy for the mass of the dark matter halos hosting galaxies. 
		In order to select halos hosting CMASS galaxies, we create a new variable $V_{new}=(1+\mathcal{N}(\sigma_{\textsc{ham}}))V_{peak}$, where $\mathcal{N}$ is a random number from a Gaussian distribution with mean 0 and standard deviation $\sigma_{\textsc{ham}}$. 
Then, we rank-order halos based on $V_{new}$ and select a number above a threshold in $V_{new}$ tuned to achieve the desired number density. In this procedure we do not distinguish between host and sub-halos, so the fraction of subhalos in the light-cone is given by the simulations. This methodology reproduces with a good agreement the small scales of both samples, with 11.4\% and 12.3\% of subhalos in the CMASS and LOWZ lightcones respectively.  

		We construct CMASS and LOWZ light-cones including all the available snapshots (40), which allows for a correct evolution of the dark matter along the line of sight. In order to reproduce the evolution of the observed clustering, we implement a SHAM model in six redshift bins using a different scatter value for each one. 
	This is due to the dependency of the scatter with the circular velocity presented in the methodology used in the construction of our mock. However, our performance allows us to select the dark matter halos hosting LRGs using their abundance and clustering signal, without any  stellar mass (or luminosity) information. \citep[see][]{Nuza2013}.

	Many studies have used SHAM models to study LRG galaxies
for different stellar mass cuts \citep[e.g.][]{Reddick2013,Tinker2017}, ensuring that 
the observed sample
is complete for the given stellar mass cuts. In these cases, the clustering is described by the number density and the intrinsic scatter between galaxies and halos. Nevertheless, the BOSS sample is not complete for all stellar masses, so we have to take it into account for the construction of our mock. Different studies have included the incompleteness of the sample using stellar mass information \citep[e.g.][]{Saito2016,Rodrigueztorres2016}; however, we include this effect without using a proxy for galaxy mass, 
by increasing the value of scatter. This allows us approximately to mimic the incompleteness, including less massive halos in the final catalogue. 
Thus, the scatter used to reproduce the clustering in each redshift is combining the effect of the intrinsic relationship between galaxies and halos and the incompleteness of the sample. Most of the values of scatter are found to vary between 0.1 and 0.24, with an exception at high redshift  ($0.57<z<0.7$) where the scatter is $0.55\times V_{peak}$. Nevertheless, this is expected, since the abundance of galaxies is lower at this redshift (the mean stellar mass is higher than in other redshifts), while the clustering signal is relatively consistent with the clustering at other redshifts. 
The scatter values used and their dependence on redshift can be seen in Table \ref{tab:lc}.
	
	\begin{table}
          \centering
          \begin{tabular}{cccc}\hline
          sample & $z$ range & $\sigma_{\textsc{ham}}$ & $\bar{n}$ ($10^{-4}$Mpc$^{-3}h^{3}$)\\ \hline \hline
          &0.16-0.24&0.26&4.10\\
          LOWZ    &0.24-0.30&0.10&3.05\\
          &0.30-0.36&0.14&3.29\\ \hline
          &0.43-0.51&0.31&3.26\\
          CMASS  &0.51-0.57&0.24&3.66\\
          &0.57-0.70&0.55&1.43\\ \hline
          \end{tabular}
          \caption{Observed number density and scatter values used to select the halos hosting CMASS galaxies in six different redshift ranges.}
          \label{tab:lc}
        \end{table}

		Additionally, we produce light-cones with different satellite fractions. This is done by applying the same scatter for subhalos and host halos, such
		as in the standard HAM, but in this case, we include a new parameter to select a certain fraction of subhalos. The fraction of satellites of the
		standard HAM is varied by $\pm10$\% and $\pm15$\%, that means, an increase of 10\% modifies the subhalo fraction of the LOWZ like sample from 12.3\%
		to 13.5\%. Note that there are additional selection effects such as mass and color dependent incompleteness that can potentially modify the
		galaxy clustering and galaxy-matter cross correlations on small scales \citep{Leauthaud2017}.
		In this work we are primarily interested in testing the variations
		on \rcc\ at scales $r_p>1\mpch$ at fixed clustering, where we expect that impact of these effects to be small on top of the variations introduced by
		varying satellite fractions.

\section{Results}\label{sec:results}
	In this section we first present the results of the analysis on mock galaxy catalogs, testing the impact of different estimators, and other
	modeling  assumptions (see also the appendices)
    and then we present the results of
    analysis on the data, using both galaxy lensing and CMB lensing. Throughout this section
	we define $S_8=\left(\frac{\sigma_8}{0.8228}\right)^{0.8}\left(\frac{\Omega_M}{0.307}\right)^{0.6}$, where the normalization factors for $\sigma_8$
	and $\Omega_M$ are chosen based on the cosmology in the mock samples. We obtain the parameter likelihoods
    using MCMC with flat priors on $A_s\in[10^{-10},10^{-8}]$ and $\Omega_m\in[0.05,0.75]$ and scale-dependent priors on \rcc. We fix other
    cosmological parameters and we do not impose
    any limits on
    $\DS_0$, as well as the parameters for the \rcc\ fitting function (eq.~\ref{eq:rcc_predict}).

	\subsection{Analysis of mock galaxy catalogs}
    	In this section we present the results from fitting the models presented in Section~\ref{sec:modeling} to the measurements from mock galaxy catalogs
        described in Section~\ref{ssec:z_mocks}

			\begin{figure*}
        		\centering
         		\includegraphics[width=0.8\linewidth]{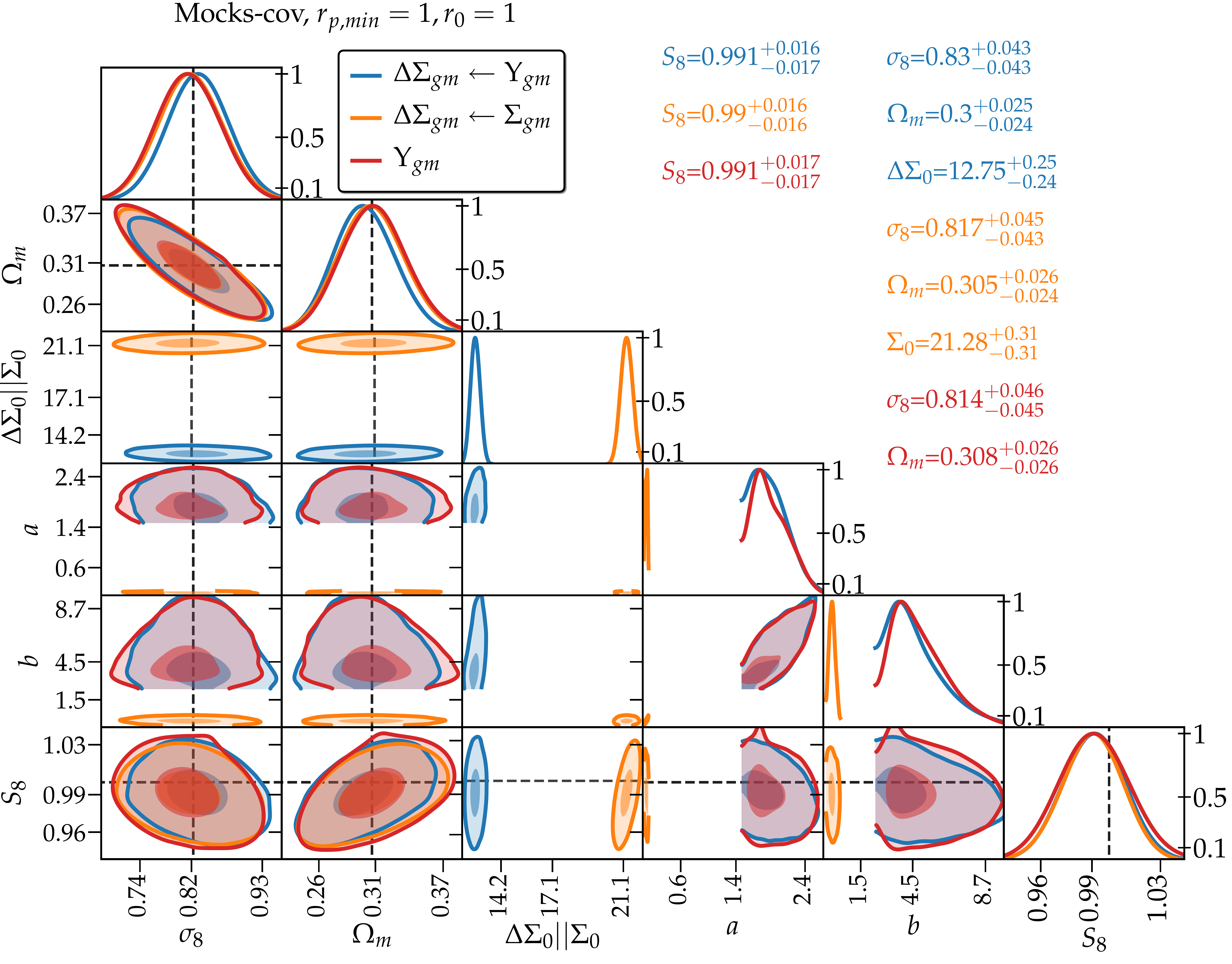}

		     \caption{Parameter constraints when fitting the correlation functions from mock galaxy catalogs using different
             estimators or modeling schemes discussed in section~\ref{sec:modeling}, using $r_0=1\mpch$. Note that we define
             $S_8=\left(\frac{\sigma_8}{0.8228}\right)^{0.8}\left(\frac{\Omega_M}{0.307}\right)^{0.6}$ and its expected value is 1.
             Blue color represents the modeling of \ugm\ as discussed in
             section~\ref{ssec:modeling_upsilon} while red and green show two different methods for modeling \DS\ by predicting \ugm\ or $\DS_{gm}$
             as discussed in section~\ref{ssec:modeling_DS}.
             These results are using
			with covariance matrix from mock catalogs (no shape noise). We applied flat priors on all parameters, with $\Omega_m\in[0.05,0.75]$ and
			$A_s\in[10^{-10},10^{-8}]$ ($A_s$ posterior is not shown, instead we show the derived parameter $\sigma_8$ based on $A_s$ values in the
            chain). There were no limits imposed on values of $a,b,\DS_0,\Sigma_0$, but we did impose flat priors on \rcc\ based on the range from
            mock catalogs \refereeTwo{which implicitly imposes cuts on $a,b$}, \referee{as discussed in section~\ref{sec:modeling}}.
		     Filled dark contours enclose the 68\% region
		     while solid lines enclose the 95\% region.
             Numbers quoted in the figures are the maximum likelihood values while
		     errors bars are from the 16 and 84 percentile.
			Vertical and horizontal lines mark the fiducial values for
		     different parameters. Note that $\Delta \Sigma_0$
                     and $\Sigma_0$ share the same panel, although they
                     are different parameters that should not be
                     compared numerically. \referee{Also, the limits on the panels are determined from the $\pm3\sigma$ spread (marginal) of 
                     the parameter values for all the chains plotted.}
		      }
		     \label{fig:mocks_estimator_comp}
    	    \end{figure*}

	    	\begin{figure}
        		\centering
         		\includegraphics[width=\columnwidth]{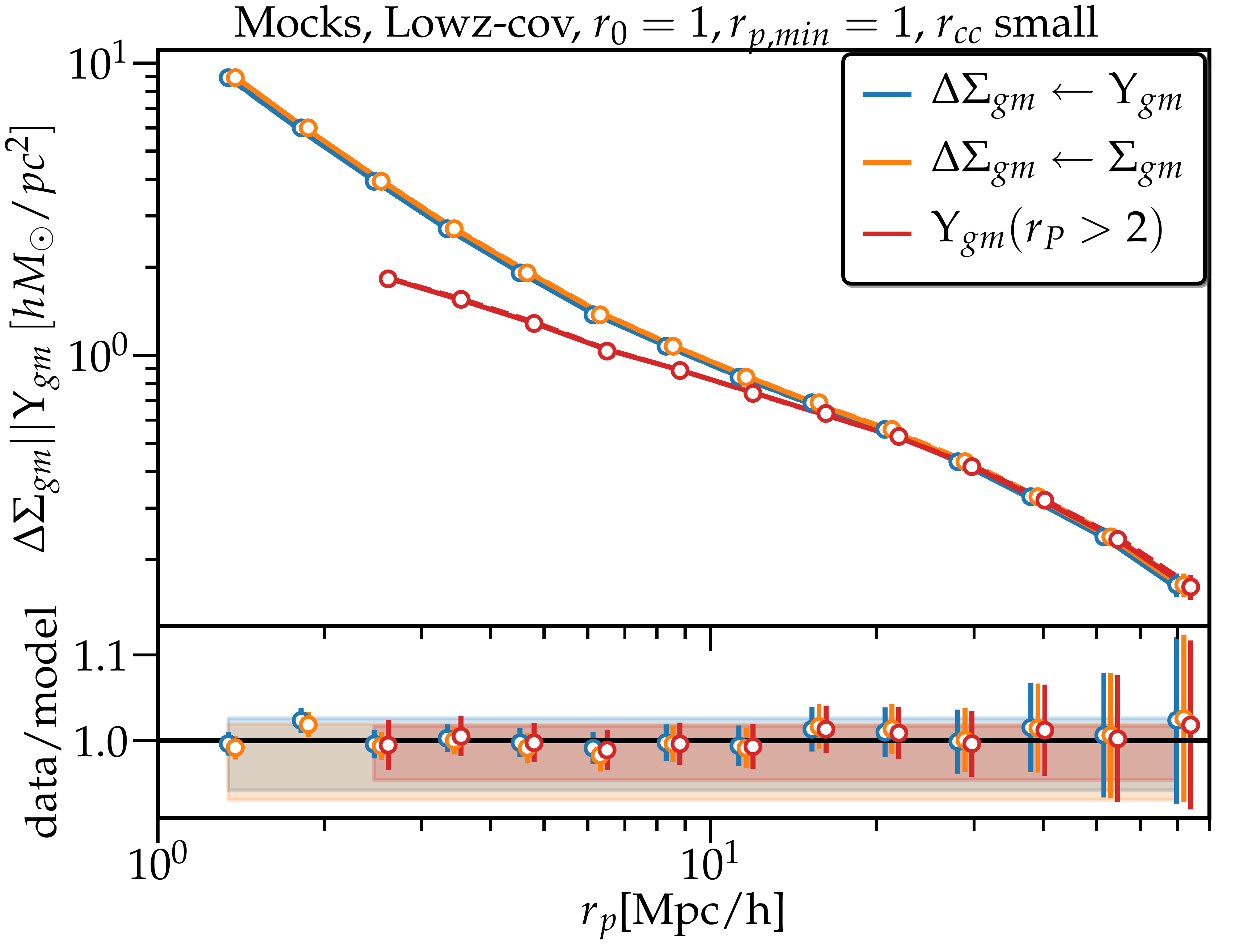}
				\caption{Upper panel: Model fits to the correlation functions from mock galaxy catalogs using best-fitting values (solid
                	lines) from Fig.~\ref{fig:mocks_estimator_comp}. Open points show the measurements
                    from mock catalogs along with the errors while the solid lines show the best fit model.
                    Dotted lines (not distinguishable form solid lines) are using the
					maximum likelihood point around the fiducial values of $\sigma_8$ and $\Omega_m$
                    (within 0.1\% of fiducial
					values). Note that the errorbars on the points are from  the covariance of mock catalogs (no shape
                    noise).
               		Bottom panel: Open points show the ratio of the data points to the model predictions.
                    The filled contours show the ratio of the best fit model to the model from the
                    fiducial (Planck) cosmology, \referee{where the contour width indicates the $1\sigma$ range on $S_8$}.
				}
	    	     \label{fig:mocks_estimator_data_comp}
		     \end{figure}

		Fig.~\ref{fig:mocks_estimator_comp} shows the results of fitting measurements in mock catalogs using different  models and using the
        jackknife covariance from the mock galaxy catalogs (no shape
		noise).
		We get consistent best fit values as well as consistent signal to noise ratio ($S/N\sim 70$) with different estimators and the results are
        consistent with the
        fiducial model.

		Fig.~\ref{fig:mocks_estimator_data_comp} shows the comparison between mock galaxy catalogs and model predictions as well as residuals between the two.
		Model
		predictions are consistent with mock catalogs within the noise and we observe no trends in the residuals. These results suggest that
		our model works well down to $\sim1\mpch$ scale, though it not very surprising since we are using the priors on \rcc\ derived from the
        mock catalogs used themselves.
 		In Appendix~\ref{appendix:mocks} we show results with a different set of mock catalogs, with different
		satellite fraction and \rcc,
		where we obtain consistent results using $r_0=2\mpch$ and slightly biased  results ($\sim1\sigma$) using $r_0=1\mpch$.  For most of the
        tests in
		this section, we will continue to use the more aggressive choice of $1\mpch$, though for the main results using the data we will show
        results with both $r_0=1\mpch$ and $2\mpch$.

		\begin{figure}
        	\centering
         	\includegraphics[width=\columnwidth]{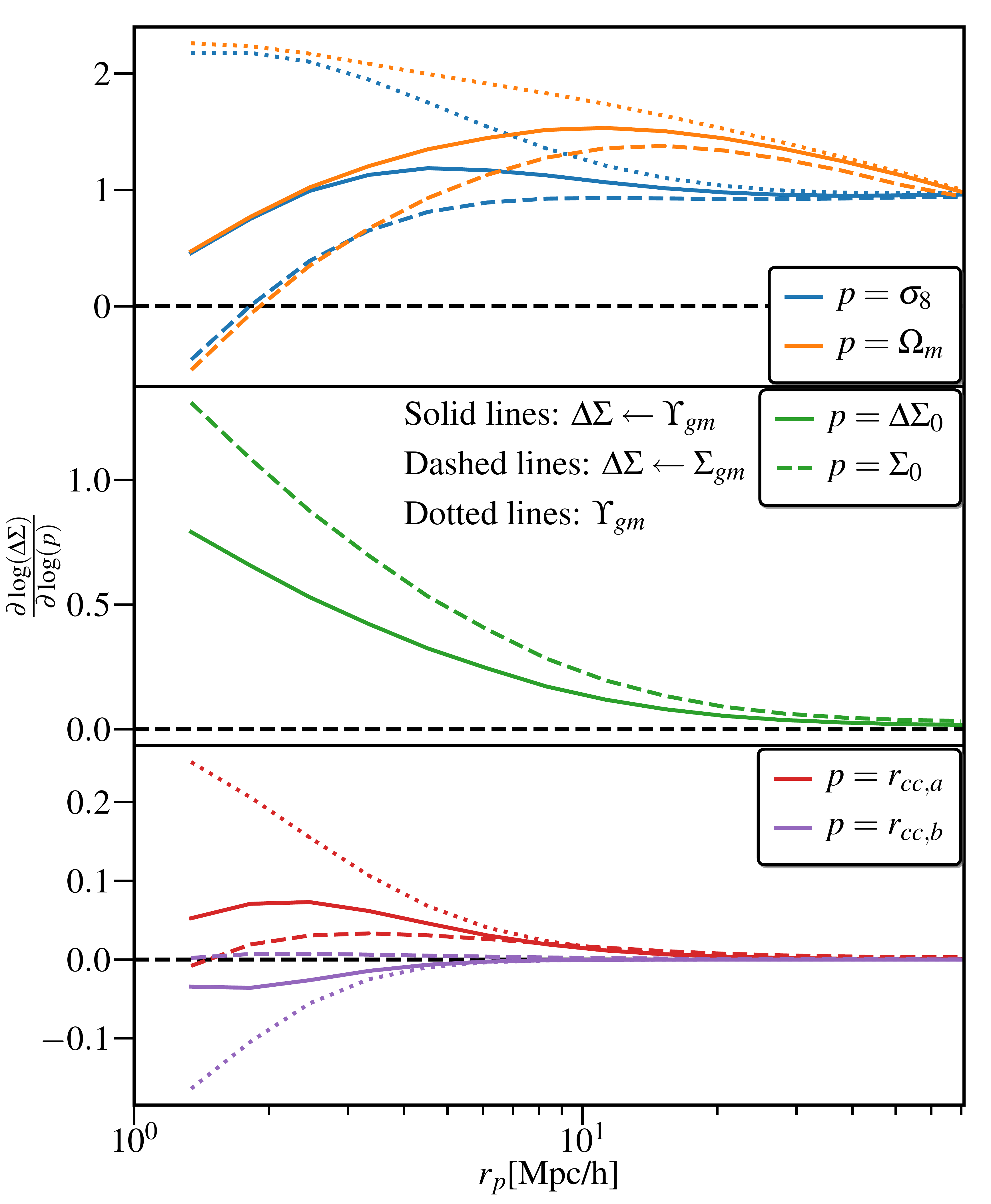}
			\caption{ Derivatives of the models with respect to $\sigma_8,\Omega_m$ (upper panel), $\DS_0$ (middle panel) and $\rcc$ parameter
				(lower panel).
				\ugm is more sensitive to cosmological parameters at smaller scales. However, when including $\DS_0$ into the model, the
				sensitivity to
				cosmological parameters decreases at small scales as the impact of $\DS_0$ increases.
				The scales we use are relatively less sensitive to the \rcc\ parameters.
				}
	    	     \label{fig:model_derivatives}
		\end{figure}

		In Fig.~\ref{fig:model_derivatives}, we show the derivatives of the different models with respect to different parameters. The
        sensitivity of the
		model to cosmological parameters increases as we go to smaller scales. However, small scales are also very sensitive to $\DS_0$ which
        suggests
		that some small-scale information is used in determining $\DS_0$. On the scales that we use, the model is relatively less
		sensitive to the $\rcc$ parameters.
		In the following sections we explore the impact of some of the choices in the analysis in more detail.
        We further show some tests based on a different set of mock galaxy catalogs in Appendix~\ref{appendix:mocks} which highlight the impact of different
        choices of $r_0$ values. In Appendix~\ref{appendix:covariance} we show that the
        different covariances (jackknife and analytical) give consistent results. In Appendix~\ref{appendix:baryons} using a toy model
        we show that baryonic effects can introduce biases of up to 2\% in \SO.
		Parameter values obtained using different
		modeling
		assumptions are shown in Table~\ref{tab:mock_params} and also on the figures showing different tests.

		\subsubsection{Impact of $\DS_0$ or $\Sigma_0$}
			\begin{figure}
        		\centering
         		\includegraphics[width=\columnwidth]{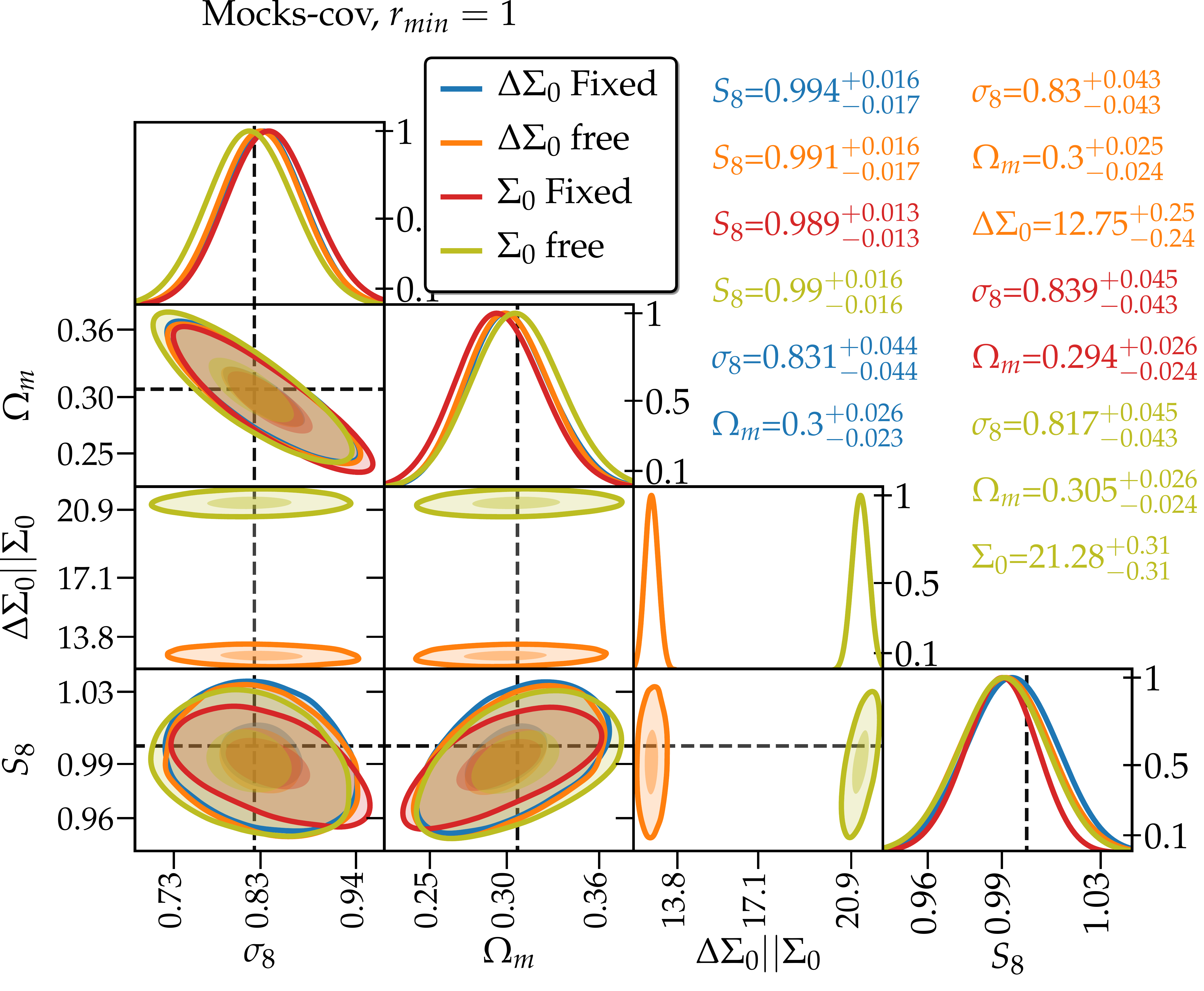} 
		         \caption{Comparison constraints obtained with fixed $\DS_0/\Sigma_0$
		         and marginalizing over $\DS_0/\Sigma_0$ for the two models used to fitting \DS.
		         }
	    	     \label{fig:mocks_DS_fix_DS0}
			\end{figure}
			In Fig.~\ref{fig:mocks_DS_fix_DS0}, we show the comparison of constraints for fixed $\DS_0$ versus marginalizing over $\DS_0$ when
			fitting \DS. Note that we do not impose any limits on $\DS_0$ in the likelihood analysis and
            when fixing $\DS_0$ we obtain the $\DS_0$ value at fixed cosmology and \rcc\ (\rcc\ is fixed to its value for the
            mock galaxy catalogs and we checked that the $\DS_0$ value obtained in this procedure is consistent with the measured value from mock catalogs).
			With fixed $\DS_0$, the S/N in \SO\ for the two different models to fit $\DS$ is different by $\sim30\%$.
			This is likely due to the projection effects onto \SO\ plane, though the two models also have different
			sensitivity to
			the $\DS_0$ as shown in Fig.~\ref{fig:model_derivatives} which can contribute to these differences. After marginalizing over $\DS_0$, the two models give consistent
			S/N. This
			is expected as even though the models are somewhat different, both models are effectively marginalizing over the $\DS_0$ as was
			shown in Eq.~\eqref{eq:sigma0_DS0}. For results shown in the paper hereafter we will always marginalize over $\DS_0$ or $\Sigma_0$,
            unless explicitly stated otherwise.

		\subsubsection{Impact of \rcc}
			\begin{figure}
            		\centering
             		\includegraphics[width=\columnwidth]{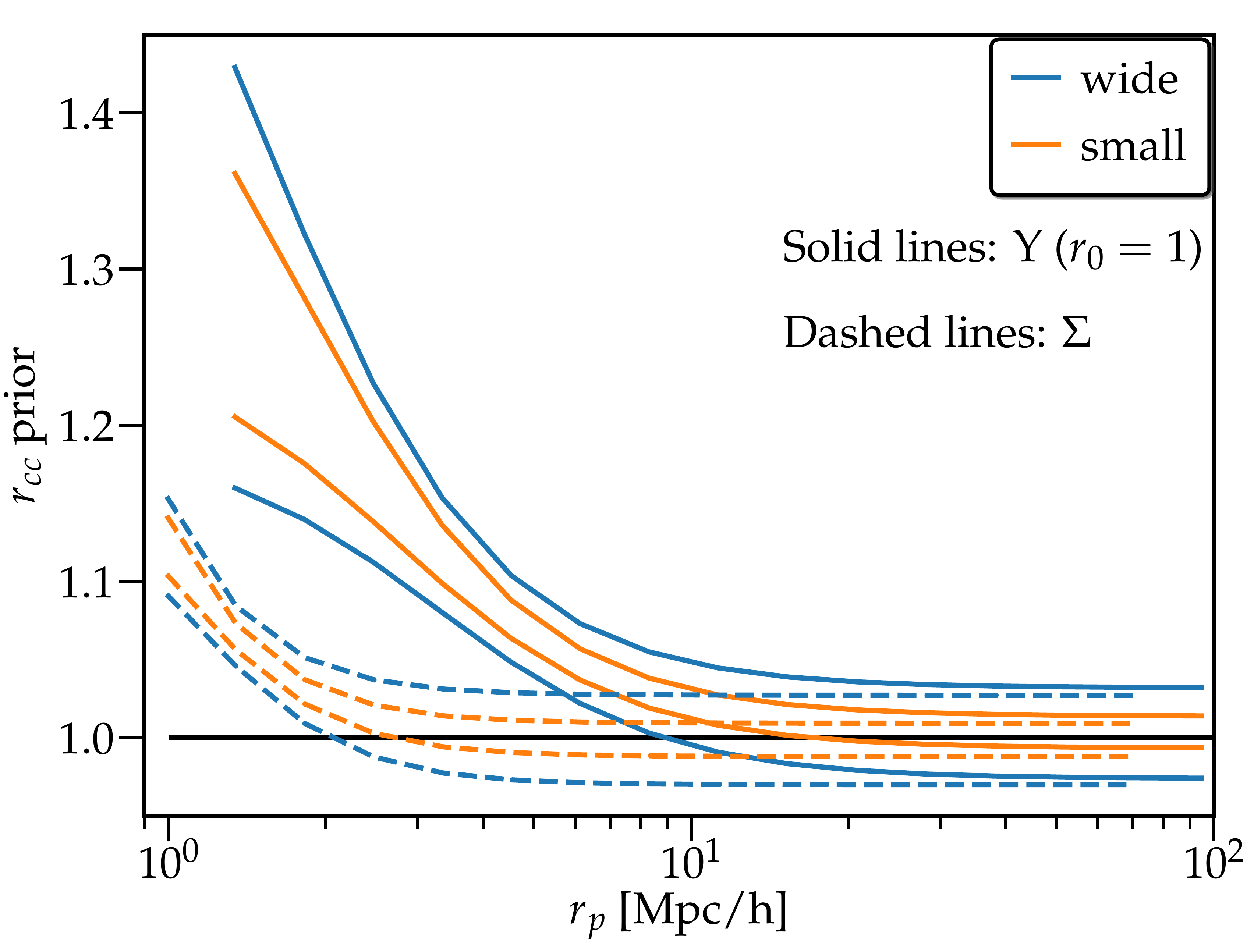}
    		         \caption{\rcc\ priors obtained from mock galaxy catalogs with different satellite fractions. `Small' priors are computed
    		         	using variations in the mean \rcc\ values while the `wide' priors account for statistical uncertainty also (variations in
    					$\rcc\pm\delta\rcc$).
    					}
    	    	     \label{fig:rcc_comp}
    	    \end{figure}
	    	In Fig.~\ref{fig:rcc_comp} we show the priors used on \rcc. Priors labeled as `small' are computed from the variations in the mean
			\rcc\ as we vary the satellite fraction in the mock catalogs , while
			the `wide' priors are computed by accounting for the noise in \rcc\ ($\delta\rcc$) as well, i.e., the variations in
			$\rcc\pm\delta\rcc$.
            `wide' priors are used to test the impact of \rcc\ priors on \SO\ and study how parameters change as we loosen the
			\rcc\ priors. Note that for our fiducial analysis we will use the `small' priors.

			\begin{figure}
        		\centering
         		\includegraphics[width=\columnwidth]{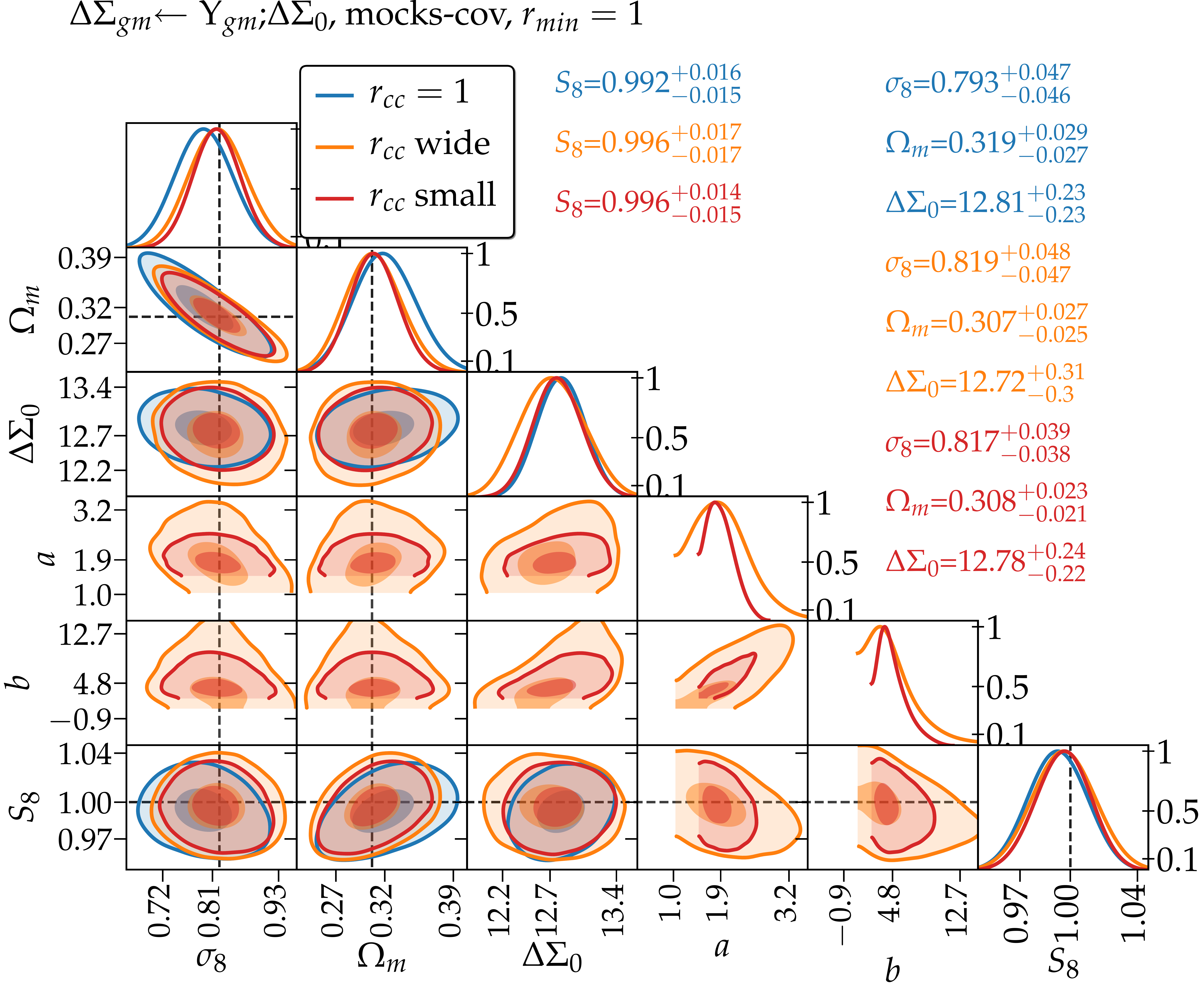}
		         \caption{ Constraints obtained using different priors on \rcc, using covariance from mock galaxy catalogs (no shape noise).
		         All the priors give results consistent with the fiducial cosmology, with $\rcc=1$ having the largest bias $\sim 0.3\sigma$.
		         The constraints using the `small' priors are tighter compared to those using wide priors by $\sim20\%$.
		         }
	    	     \label{fig:DS_rcc_comp}
    	    \end{figure}

	    In Fig.~\ref{fig:DS_rcc_comp} we show the impact of various choices of \rcc\ priors on the constraints on \SO. Using the small
	    priors on \rcc\
	    only makes a small impact on \SO, and both the signal and noise are close to the values obtained by fixing \rcc\ to the correct value
	    for the
	    mock catalogs . Relaxing the \rcc\ priors to `wide' increases the noise though the best fit values are very similar to the `small' case.
	    Removing the
	    priors on \rcc\ further, only forcing \rcc\ to be 1 on largest scales, increases the noise further by 20-30\%
	    though the best-fitting \SO\ values are unchanged (not shown).
	    We also test the results by fixing \rcc\ to 1 on
	    all scales.
	    This leads to a small bias in the best-fitting values of \SO\ though the results are consistent with the fiducial
        cosmology (with the bias in \SO\ being $\sim0.3\sigma$).
	    This is because $\DS_0$ absorbs most of the effects of the incorrect \rcc\ in this case, and since \SO\ is degenerate with $\DS_0$,
	    setting \rcc\ to 1 also increases the noise in \SO\ (relative to \rcc\ fixed at correct values).

		\begin{table}
  		   \centering
 		      \begin{tabular}{|c|c|c|c}\hline
    			Estimator & Model & $r_{cc}$ prior & $\SO-1$ \\ \hline \hline
				$\Upsilon_{gm}$ & $\Upsilon_{gm}$ & small & $-0.01^{+0.016}_{-0.016}$ \\ \hline
$\Upsilon_{gm}$ & $\Upsilon_{gm}$ & Fixed & $-0.01^{+0.016}_{-0.015}$ \\ \hline
$\Upsilon_{gm}$ & $\Upsilon_{gm}$ & wide & $-0.012^{+0.017}_{-0.015}$ \\ \hline
$\Upsilon_{gm}$ & $\Upsilon_{gm}$ & 1 & $-0.012^{+0.017}_{-0.016}$ \\ \hline
				$\Delta\Sigma_{gm}$ & $\Upsilon_{gm}$  & small & $-0.009^{+0.016}_{-0.017}$ \\ \hline
$\Delta\Sigma_{gm}$ & $\Upsilon_{gm}$  & 1 & $-0.008^{+0.016}_{-0.015}$ \\ \hline
$\Delta\Sigma_{gm}$ & $\Upsilon_{gm}$  & wide & $-0.004^{+0.017}_{-0.017}$ \\ \hline
$\Delta\Sigma_{gm}$ & $\Upsilon_{gm}$ & fixed & $-0.012^{+0.017}_{-0.016}$ \\ \hline
$\Delta\Sigma_{gm}$ & $\Sigma_{gm}$ & small & $-0.01^{+0.016}_{-0.016}$ \\ \hline
$\Delta\Sigma_{gm}$ & $\Sigma_{gm}$ & 1 & $-0.012^{+0.017}_{-0.016}$ \\ \hline
$\Delta\Sigma_{gm}$ & $\Sigma_{gm}$ & fixed & $-0.012^{+0.017}_{-0.016}$ \\ \hline
					\end{tabular}
		   	\caption{Table showing the constraints obtained on $\SO$ \referee{in an analysis of the mock catalogs, 
		   			using different models and \rcc\ prior choices}. Constraints using $\DS_\text{gm}$ are marginalized
                    over $\DS_0$ or $\Sigma_0$.
			}
			\label{tab:mock_params}
		\end{table}

		\subsubsection{Effects of theory computation}
			\cite{Troxel2018} pointed out that for coarse bins the scale at which the theoretical prediction is computed can have an impact on
            the constraints on cosmological
			parameters. If these effects become important, then the correct thing to do is to bin the theory predictions
			using the same weights as data (or use narrower bins). In
			Appendix~\ref{appendix:binning_effect} we
			show that for our results, computing the theoretical prediction at the center of the bin (arithmetic mean) has negligible impact on
            the cosmological
			parameters compared to the binned theoretical prediction. 
			Given that we are not sensitive to this choice, for simplicity, all the results shown in the rest of the paper will use the
            theoretical prediction computed at the central bin value.

	\subsection{Analysis of real data}
		In this section we apply our methodology to the galaxy clustering and galaxy-lensing cross correlations presented in
		\cite{Singh2017cmb}. \referee{We use the same pipeline as for the analysis of the mock catalogs, and vary $A_s\in[10^{-10},10^{-8}]$, $\Omega\in[0.05,0.75]$ 
		and $r_{cc}$ with scale-dependent priors derived from the mock catalogs. \SO\ is defined as 
		$S_8=\left(\frac{\sigma_8}{0.8228}\right)^{0.8}\left(\frac{\Omega_M}{0.307}\right)^{0.6}$}.
		\begin{figure*}
			\begin{subfigure}{\columnwidth}
        		\centering
         		\includegraphics[width=\columnwidth]{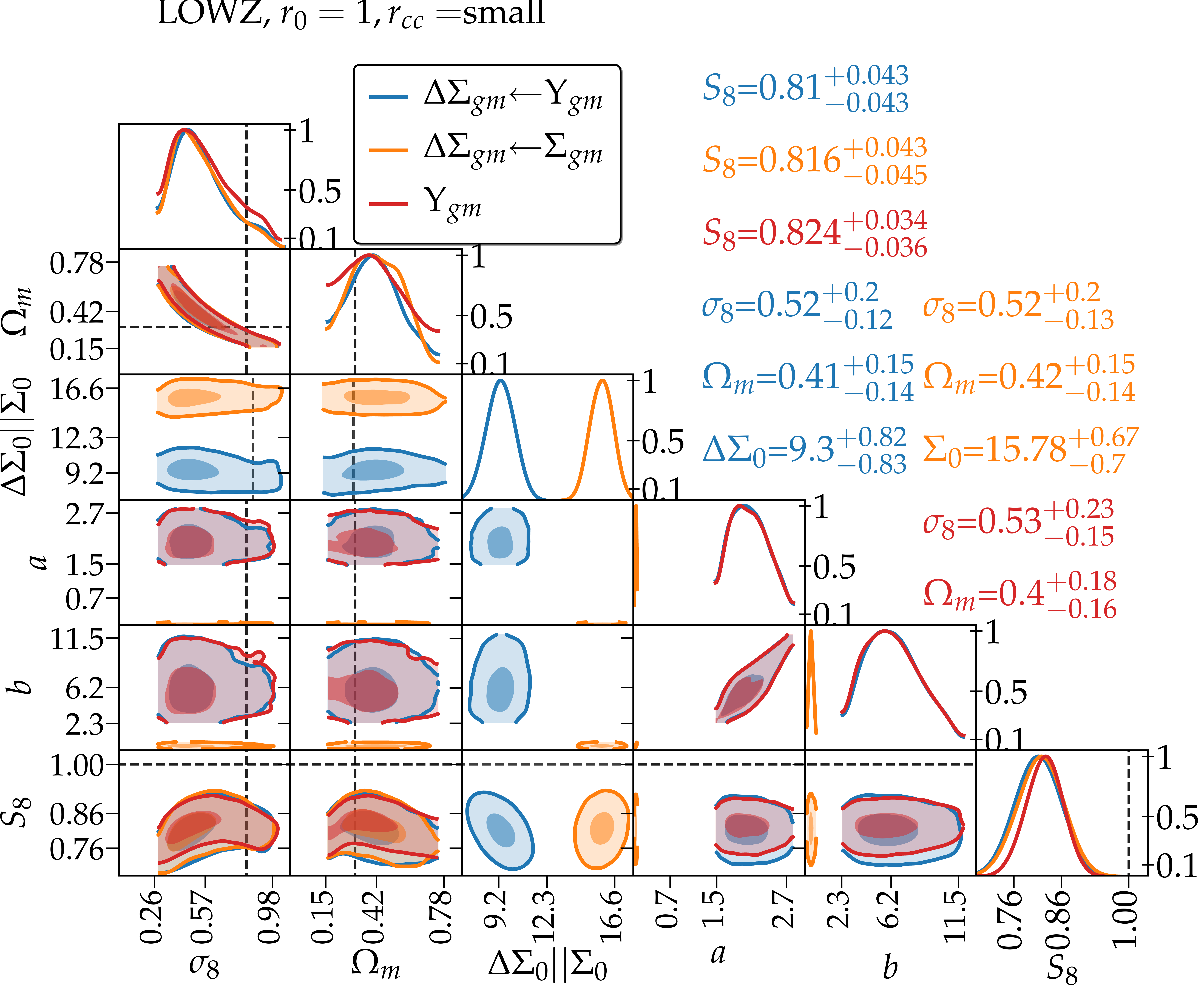}
				\caption{}
	    	     \label{fig:lowz_estimator_comparison}
		     \end{subfigure}\hfill
		     \begin{subfigure}{\columnwidth}
        		\centering
		         \includegraphics[width=\columnwidth]{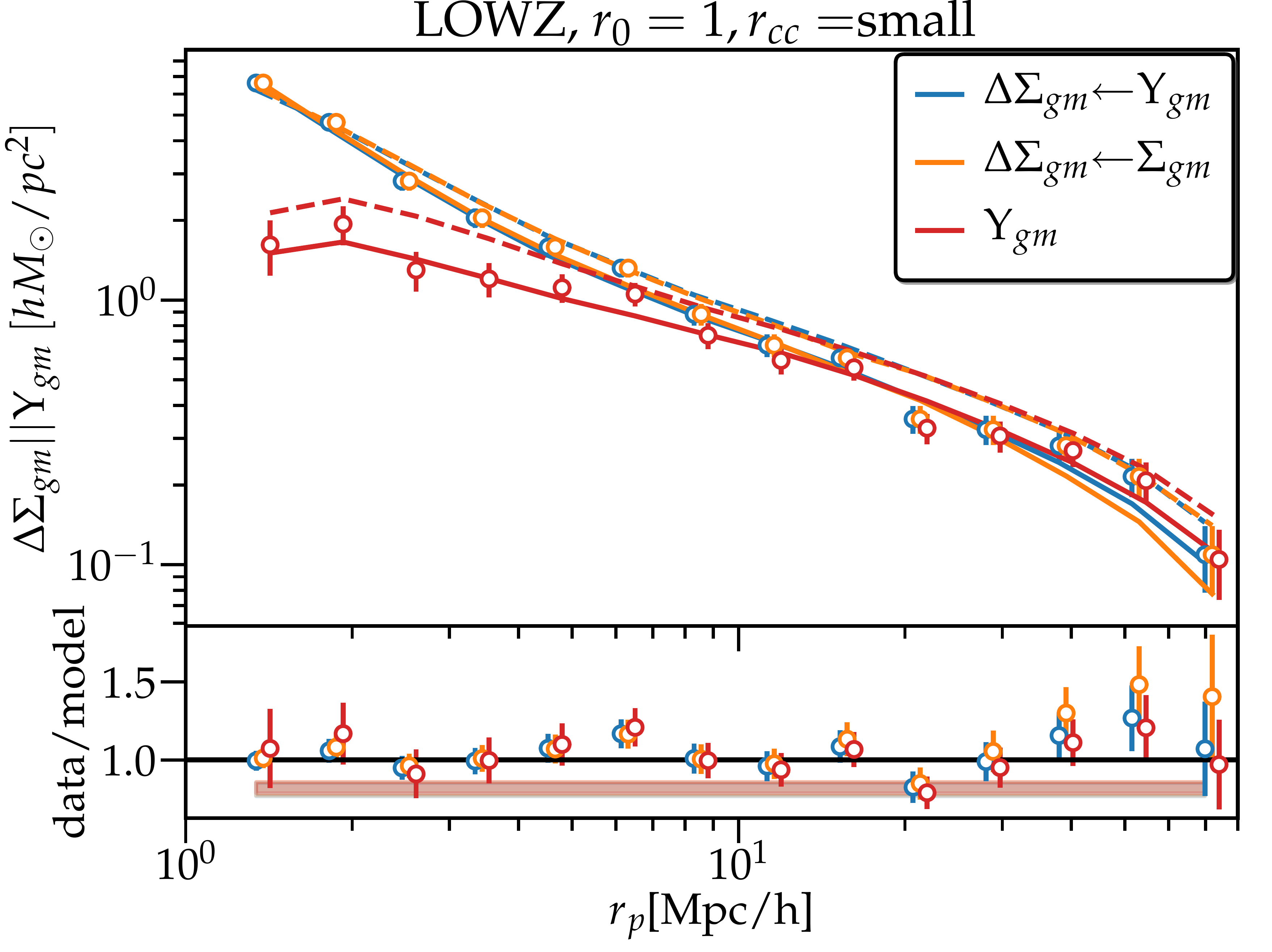}
				\caption{}
	    	     \label{fig:lowz_cosmology_model_dat_comp}
		     \end{subfigure}\hfill
                     \caption{a) Cosmological parameter constraints obtained using LOWZ data with
                     	different estimators. We obtain consistent results with
                       different estimators though the results are discrepant with the fiducial cosmology
                       at more than $3\sigma$ (stat), though
                       the impact  of systematic uncertainties (observational and theoretical)
                       are similar to the statistical errors.
						Vertical dashed lines show the Planck
                        best-fit values.  b) Model fits to the LOWZ data
                       using best-fitting values (solid lines). Dotted lines are using
                       the most likely values of nuisance parameters around the fiducial values of
                       $\sigma_8$ and $\Omega_m$. The lower panel
						show the ratio of data and best-fitting model. The filled contour shows the
                        $1\sigma$ range of the best-fitting \SO\ relative to
                       the fiducial value of \SO.
                              }
		     \label{fig:lowz_cosmology}
    	    \end{figure*}

		    In Fig.~\ref{fig:lowz_estimator_comparison} we show the cosmological parameter constraints obtained using the BOSS LOWZ sample and galaxy lensing using
		    SDSS data. While the results using different estimators are consistent, there is significant tension with the \SO\ from the Planck \lcdm\ model due to the
		    data being lower than the predictions from the best-fitting Planck cosmology by $3-4\sigma$.
		    Comparison between the data and the model is shown
		    in Fig.~\ref{fig:lowz_cosmology_model_dat_comp}.
		    The best-fitting model agrees well with the data, with no significant deviations; note that the points at large $r_p$ are correlated.
            The model with the Planck best fitting
		    cosmology,
            shown by dotted lines, predicts significantly larger $\DS$ and $\ugm$ relative to the measurements from the data.

		     The tension observed in Fig.~\ref{fig:lowz_estimator_comparison}
		     is consistent with the tension observed in the $E_G$ measurements in \cite{Singh2018eg},
		     though the significance is larger here as the noise in $E_G$ has a significant contribution from the growth rate measurements, which
             are not included in this work. As was discussed in
		     \cite{Singh2018eg}, there are potential systematic effects in shear calibration and photometric redshifts at the $\sim6\%$ level, which
             is comparable
		     to the statistical uncertainties in our results. Even after adding $6\%$ additional systematic uncertainty, we observe $2-3\sigma$
             tension with
		     the Planck cosmology.
		     \refereeTwo{Possible explanations for the residual tension include real tension with Planck due to physics beyond
					$\Lambda$CDM, a statistical fluctuation, or residual non-linear modeling uncertainty (though our analysis of the mock
					catalogs suggests it is a small effect).}

		     The S/N of the cosmological parameter constraints in our results is lower than that predicted by \cite{Wibking2017}, by $\sim 50\%$.
		     This is primarily due to the
		     effects of lower signal relative to the predictions with the Planck cosmology as used by \cite{Wibking2017}
             (since the covariances are very similar in the two studies) and the Hartlap correction  factor when using the jackknife covariance.
 			These effects
		     account for most of the differences ($30-40\%$) while the remaining differences can arise from the different modeling assumptions as
             well
		     as the inherent inaccuracies of the Fisher forecasts (they only provide a lower bound on uncertainty in general),
			making a detailed quantitative comparison difficult.

		     \begin{figure*}
                       \centering
                       \includegraphics[width=2\columnwidth]{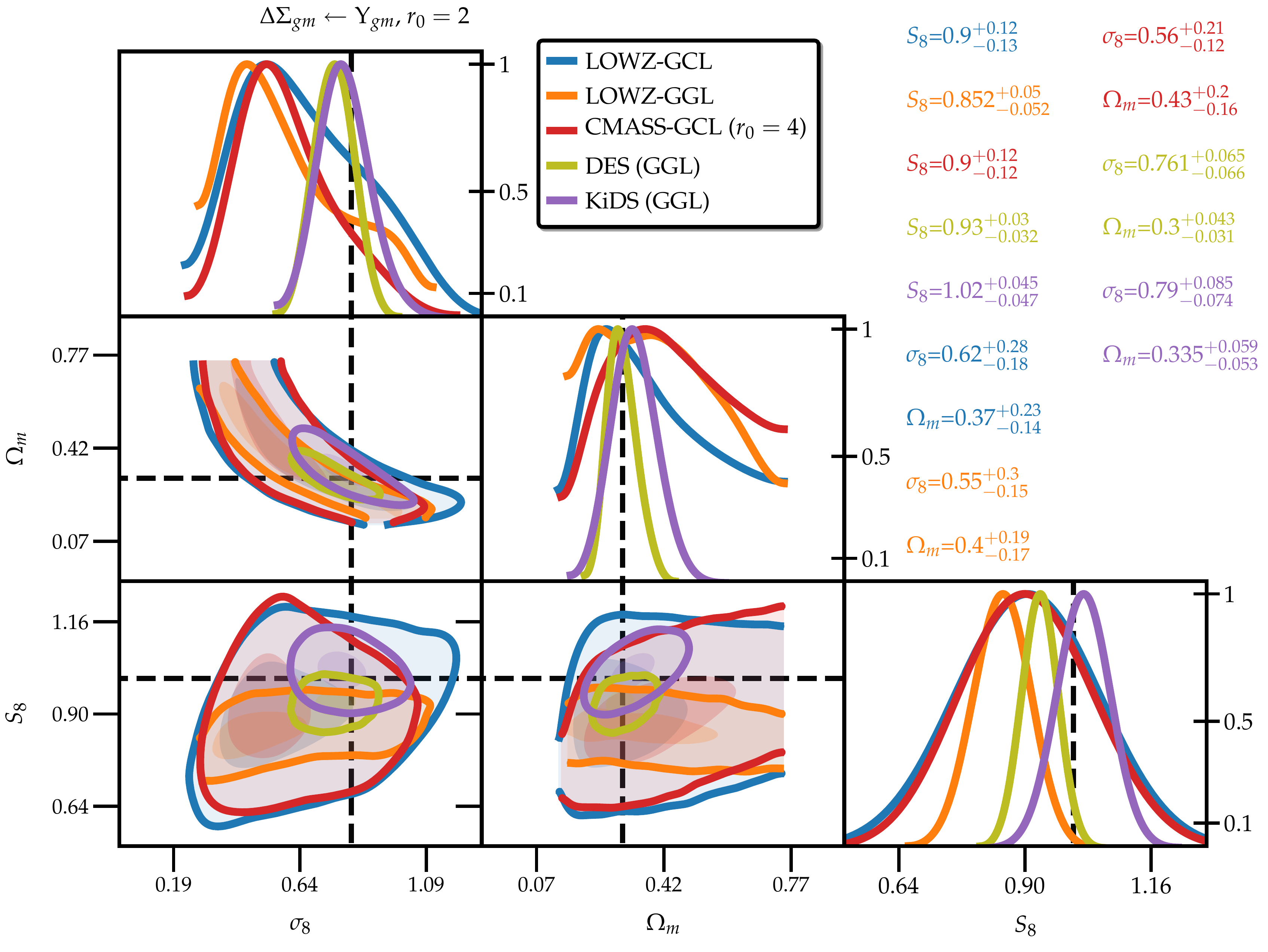} 
                       \caption{ Comparison of constraints obtained using BOSS $\times$ \referee{CMB lensing (GCL) and galaxy lensing (GGL)},
                        along with published galaxy
                       clustering+galaxy-galaxy
                       lensing results from \protect{\citealt{DES2017comb}}  and {\citealt{Uitert2018}} (KiDS$\times$GAMA). }
	    	     \label{fig:cmb_ggl_comp}
		     \end{figure*}

		 	 Fig.~\ref{fig:cmb_ggl_comp} shows the results from cross correlations between the BOSS CMASS and LOWZ samples and Planck CMB lensing. For
			 the
			 CMASS sample, we use a larger $r_0=4\mpch$, to avoid the scales affected by the Planck beam \citep[see discussion in][]{Singh2017cmb}.
			 CMB lensing also prefers a low amplitude for \SO, though the noise in these measurements is a factor of 2 larger than in the galaxy
			 lensing and
			 thus are consistent with both the galaxy lensing and the Planck \lcdm\ model at the $\sim 1\sigma$ level. In Fig.~\ref{fig:cmb_ggl_comp}
			 we also compare our results with the published results using galaxy clustering and galaxy lensing from KiDS \citep{Uitert2018} and DES
			 collaborations \citep{DES2017comb}, finding good consistency among all the results (at 95\% confidence level).

		     \begin{figure}
        		\centering
			     \includegraphics[width=\columnwidth]{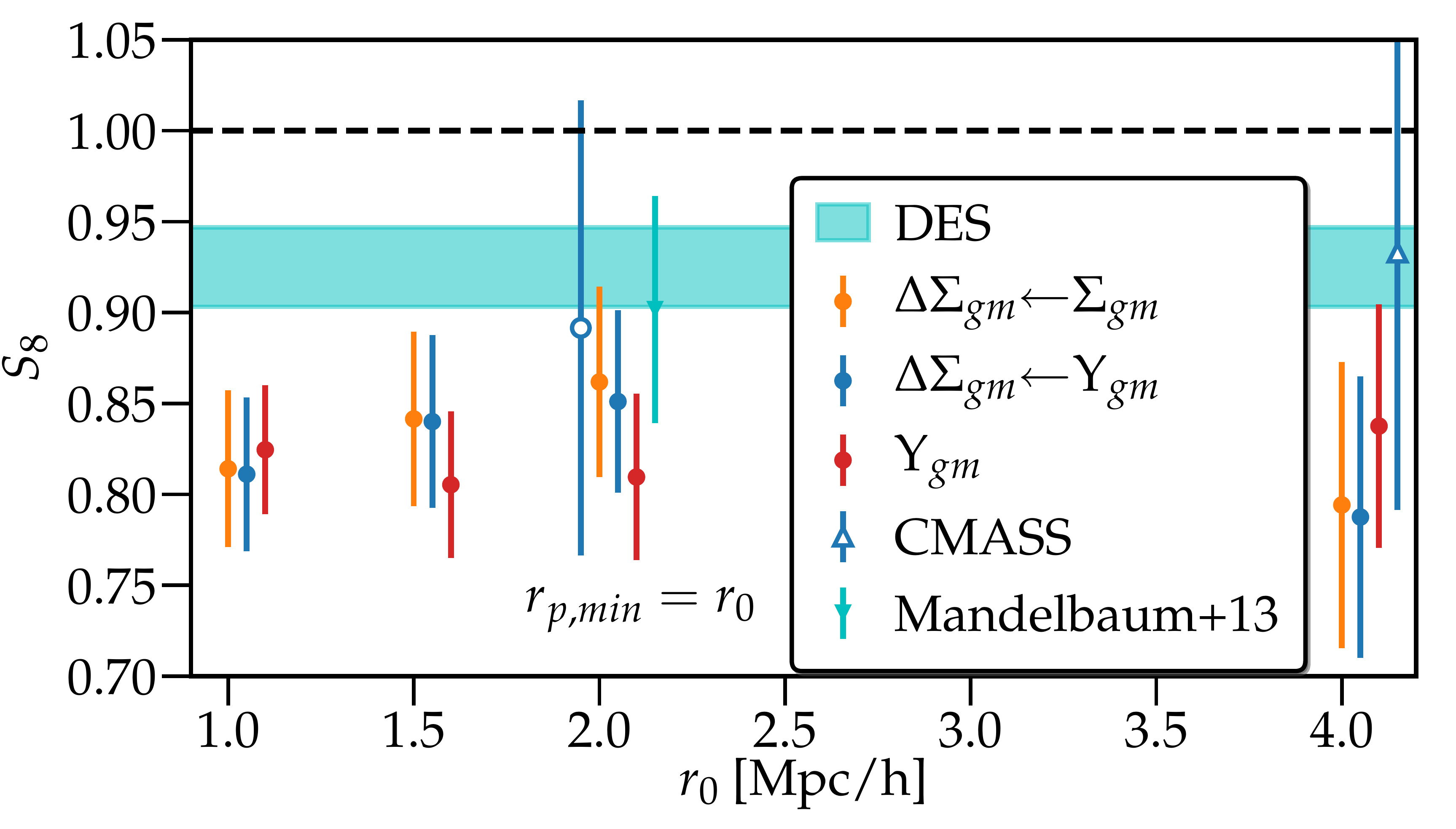}\\
				\includegraphics[width=\columnwidth]{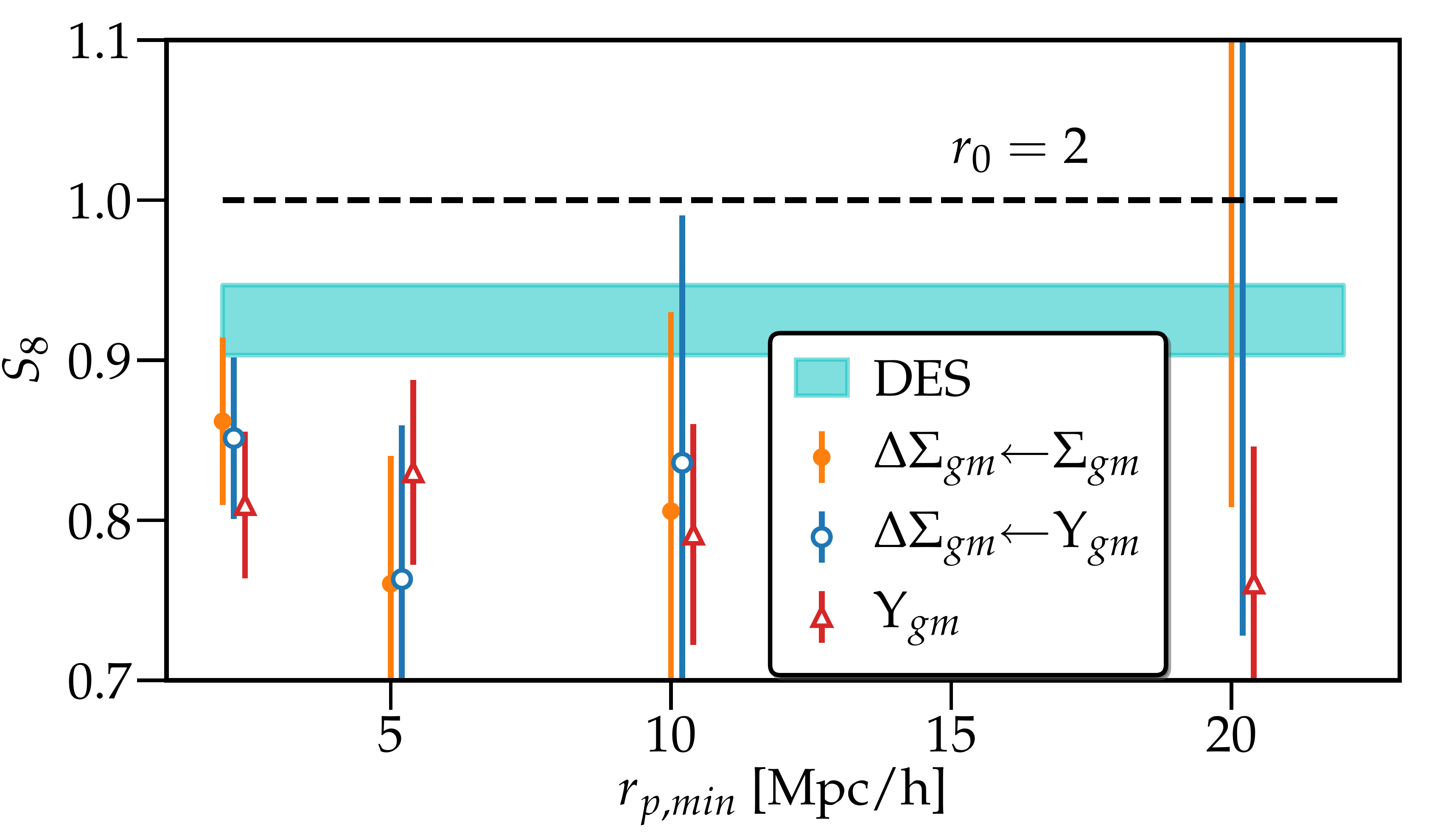}
				\caption{
						Upper panel: \SO\ as function of
						different $r_0$. We also show the results from DES \protect\citep{DES2017comb}
						and \protect\cite{Mandelbaum2013}, where we computed the central value of \SO\ using the best-fitting values of $\Omega_m$
                        and $\sigma_8$
						while the uncertainties shown are computed using relative uncertainties on their values of \SO\ (different papers have
                        slightly different definition of \SO.).
						Lower panel: \SO\ as function of different choices of $r_\text{p,min}$
                                                at fixed $r_0=2\mpch$. As the $r_\text{p,min}$ increases, the constraints on
                                                $\DS_0$ weaken, and due to degeneracy between $\DS_0$ and \SO, the
                                                constraints on \SO\ also weaken.  The \SO\ value increases at larger
                                                $r_{p,min}$ because the $\DS_0$ takes lower values ($\DS_0\sim-28\pm35$
                                                for $r_{p,min}=20$).  In both panels, points with circular markers are
                                                using the LOWZ sample, while the point with the triangular marker is using the
                                                CMASS sample. Closed markers are using galaxy lensing while open
                                                markers are using CMB lensing, and different colors mark the different
                                                estimators.  }
	    	     \label{fig:CMB_GL_comp}
		     \end{figure}
		The choice of scale used in Fig.~\ref{fig:lowz_cosmology} is fairly aggressive, as we push the analysis into non-linear scales, although the
			results of the analysis in mock galaxy catalogs is consistent with the expectation. We test the impact of the choice of scale in
			Fig.~\ref{fig:CMB_GL_comp} by choosing
			larger $r_0$ values, as well as restricting the analysis to larger scales for fixed $r_0=2\mpch$. While the errors increase with the more
			conservative choice of $r_0$ and $r_{p,min}$, the best-fitting values are consistent across the different choices of scales and lower
            compared with
			the Planck \lcdm\ model.
			This is consistent with the observation from Fig.~\ref{fig:lowz_cosmology_model_dat_comp}, where the  data is
			discrepant with the Planck model at all scales. We choose relatively
            conservative values of
            $r_0=2\mpch$ and $r_{p,min}=2\mpch$ for our fiducial results in the following
            when comparing with
            other results in the literature.

			For comparison, in  Fig.~\ref{fig:CMB_GL_comp} we also plot the \SO\ values from recent DES weak lensing results \citep{DES2017comb}
			and also from \cite{Mandelbaum2013}, who used the same lensing source catalog with a brighter lens sample across the same redshift
			range.
			The DES best-fitting cosmological parameter combination, $\sigma_8\left(\frac{\Omega_M}{0.3}\right)^{0.5}
			=0.783^{+0.021}_{-0.025}$, is consistent with our results at the $<2\sigma$ level (DES $\SO=0.925\pm0.021$ ).
			The results of \cite{Mandelbaum2013}, $\sigma_8\left(\frac{\Omega_M}{0.25}\right)^{0.57}=0.83\pm0.05$ (after rescaling by
			$1+m_\gamma\sim1.04$) are also consistent with our results at
			$\sim1\sigma$ level.

			\begin{figure*}
        		\centering
         		\includegraphics[width=2\columnwidth]{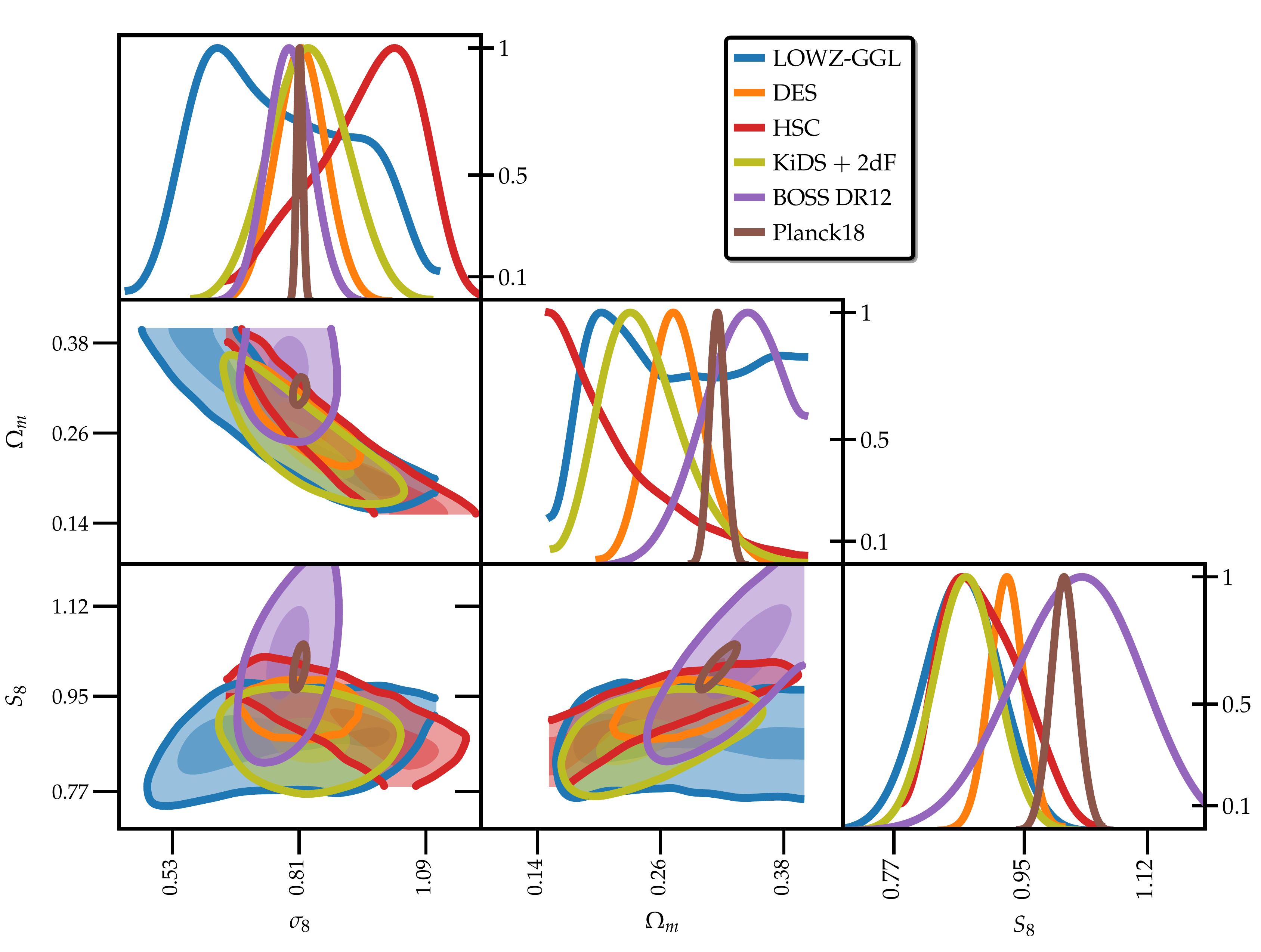}
				\caption{Comparison of Planck 2018 cosmological parameters with results obtained using galaxy-galaxy lensing from
                SDSS$\times$BOSS (this work, $r_0=2\mpch$),
                KiDS cosmic shear+KiDS$\times$2dFLens+KiDS$\times$BOSS
				(labeled KiDS$+$2dF) $\SO=0.87\pm0.04$ \protect{\citep{Joudaki2017}}, DES 3X2 analysis, $\SO=0.925\pm0.021$
				\protect{\citep[cosmic shear+galaxy-galaxy
				lensing+galaxy clustering][]{DES2017comb}} and HSC cosmic shear analysis, $\SO=0.89\pm0.05$ \citep{Hikage2018}.
				Posteriors from different datasets overlap at 95\% level, with slight preference for
				higher \SO\ in Planck posterior being driven by Planck temperature data \protect{\citep[see table 2 in][]{Planck2018}}.
				}
	    	     \label{fig:sdss_kids_planck}
		     \end{figure*}
		In Fig.~\ref{fig:sdss_kids_planck}, we show the marginalized constraints on $\sigma_8,\Omega_m$ from Planck 2018 along with some recent
		large scale structure (LSS) analysis including lensing measurements from this work,
		DES \citep{DES2017comb}, KiDS \citep{Joudaki2017} as well as
		BAO+RSD measurements from BOSS-DR12
		analysis \citep[][chains for RSD+BAO only analysis provided by Shadab Alam]{Boss2016combined}.
		The results obtained by various LSS analysis are consistent with each other though lensing measurements are somewhat discrepant with the 
		marginal constraints obtained from Planck results. Planck results prefer slightly higher \SO\ which is driven by the
		slight preference for larger \SO\ in Planck temperature data \citep[see table~2 in ][]{Planck2018}.
		In addition to systematic biases, there are physical mechanisms, eg. impact of massive neutrinos, modified gravity or dark energy models beyond 
		$\Lambda$ which can potentially reduce some of the tension. An analysis of these extensions to the standard \lcdm\ model by
		\cite{DES2018extended} though revealed no significant evidence in favor of these models with the results being
		 limited by the noise in the lensing measurements.

		\begin{table*}
  		   \centering
 		      \begin{tabular}{|c|c|c|c|c|c}\hline
    			Estimator & Model & $r_{cc}$ prior & \SO\ & $\chi^2_\text{best-fit}$ & $\chi^2_\text{fid}$\\ \hline \hline
					$\Upsilon_{gm}$ &   & small & $0.823^{+0.034}_{-0.035}$& $14.79$& $52.52$ \\ \hline
					$\Upsilon_{gm}$ &   & 1 & $0.82^{+0.035}_{-0.037}$& $15.08$& $53.78$ \\ \hline
					$\Upsilon_{gm}$ &   & wide & $0.824^{+0.035}_{-0.037}$& $13.74$& $52.74$ \\ \hline
					$\Delta\Sigma_{gm}$ & $\Upsilon_{gm}$  & small & $0.812^{+0.042}_{-0.044}$& $14.57$& $38.66$ \\ \hline
					$\Delta\Sigma_{gm}$ & $\Upsilon_{gm}$  & 1 & $0.807^{+0.044}_{-0.044}$& $15.01$& $38.98$ \\ \hline
					$\Delta\Sigma_{gm}$ & $\Upsilon_{gm}$  & wide & $0.808^{+0.042}_{-0.043}$& $13.83$& $38.71$ \\ \hline
					$\Delta\Sigma_{gm}$ th-cov & $\Upsilon_{gm}$  & small & $0.84^{+0.038}_{-0.04}$& $13.0$& $32.81$ \\ \hline
					$\Delta\Sigma_{gm}$ th-cov & $\Upsilon_{gm}$  & wide & $0.837^{+0.038}_{-0.039}$& $12.84$& $32.13$ \\ \hline
					$\Delta\Sigma_{gm}$ & $\Sigma_{gm}$ & small & $0.816^{+0.043}_{-0.045}$& $16.97$& $35.02$ \\ \hline
				\end{tabular}
		   \caption{ Constraints obtained on cosmological parameters ($S_8=\left(\frac{\sigma_8}{0.8228}\right)^{0.8}\left(\frac{\Omega_M}{0.307}\right)^{0.6}$) using the galaxy-galaxy lensing results (LOWZ$\times$SDSS), with
			 $r_0=1\mpch$. `th-cov' denotes the results obtained using analytical covariance matrix.
		   $\chi^2_\text{best-fit}$
			 shows the minimum $\chi^2$ from
		   the values within the MCMC chains (we use 14 bins in these fits) while $\chi^2_\text{fid}$ show the
			 minimum $\chi^2$ within the chains when $\sigma_8-\Omega_m$ values
		   are chosen to be within $1\%$ of the Planck values. In case of $\DS$, we do not impose any constraints
			 on $\DS_0$ or $\Sigma_0$ when obtaining $\chi^2_\text{fid}$, which results in better fit on
			 small scales and hence lower $\chi^2_\text{fid}$ (note that small scales also have higher S/N which
			 gives them larger weight in $\chi^2$).
					}
			\label{tab:data_params}
		\end{table*}

		\begin{table*}
  		   \centering
 		      \begin{tabular}{|c|c|c|c|c|c}\hline
    			Estimator & Model & $r_{cc}$ prior & \SO\ & $\chi^2_\text{best-fit}$ & $\chi^2_\text{fid}$\\ \hline \hline
					$\Upsilon_{gm}$ &   & small & $0.81^{+0.044}_{-0.046}$& $13.73$& $32.97$ \\ \hline
$\Upsilon_{gm}$ &   & 1 & $0.809^{+0.046}_{-0.046}$& $14.06$& $32.21$ \\ \hline
$\Upsilon_{gm}$ &   & wide & $0.799^{+0.052}_{-0.051}$& $11.57$& $32.73$ \\ \hline
$\pmb{\Delta\Sigma_{gm}}$ & $\Upsilon_{gm}$ & small & $0.852^{+0.05}_{-0.052}$& $10.99$& $18.98$ \\ \hline
$\Delta\Sigma_{gm}$ & $\Upsilon_{gm}$  & 1 & $0.854^{+0.051}_{-0.051}$& $10.87$& $19.06$ \\ \hline
$\Delta\Sigma_{gm}$ & $\Upsilon_{gm}$  & wide & $0.848^{+0.057}_{-0.056}$& $11.1$& $19.55$ \\ \hline
				$\Delta\Sigma_{gm}$ th-cov & $\Upsilon_{gm}$  & small & $0.849^{+0.046}_{-0.047}$& $10.92$& $21.73$ \\ \hline
$\Delta\Sigma_{gm}$ th-cov & $\Upsilon_{gm}$  & wide & $0.842^{+0.053}_{-0.054}$& $10.81$& $21.58$ \\ \hline
	$\pmb{\Delta\Sigma_{gm}}$- GCL & $\Upsilon_{gm}$  & small & $0.9^{+0.12}_{-0.13}$& $3.87$& $4.84$ \\ \hline
				$\Upsilon_{gm}$- GCL &  & small & $0.88^{+0.12}_{-0.12}$& $3.62$& $4.64$ \\ \hline
			$\pmb{\Delta\Sigma_{gm}}$ (CMASS$\times$GCL) & $\Upsilon_{gm}$  & small & $0.93^{+0.15}_{-0.13}$& $4.61$& $5.37$ \\ \hline
				$\Upsilon_{gm}$ & (CMASS$\times$GCL) & small & $0.92^{+0.13}_{-0.11}$& $4.14$& $5.15$ \\ \hline
		      \end{tabular}
		   \caption{Same as table~\ref{tab:data_params}, now showing fits with $r_0=2\mpch$ (12 bins). This table presents
            the fiducial results for our analysis, with the values in bold $\pmb{\DS_{gm}}$ showing the values quoted in
            abstract (we combined the results for CMB lensing assuming LOWZ and CMASS results are independent). Last four rows show the results with CMB lensing (GCL). $r_0=4\mpch$ for
            CMASS$\times$CMB-lensing to avoid the scales below the resolution limit of Planck CMB lensing maps.
					}
			\label{tab:data_params2}
		\end{table*}

\section{Conclusions}
\label{sec:conclusions}
	We have presented results using a new method for joint analysis of galaxy clustering and galaxy lensing
	cross correlations. Since for the current
	data, galaxy clustering has much better signal to noise compared to the galaxy-lensing cross correlations,
	we showed that galaxy clustering
	measurements can be used as part of the model itself, which eliminates
	the need for complicated galaxy-bias modeling. We still need to model the
	non-linear cross correlation coefficients, \rcc, which we do using a simple parametric model with priors taken from mock catalogs .

	We test for the impact of various uncertainties in our theoretical model. In our model, the impact of non-linearities in \DS\ get absorbed in
    the nuisance
	parameter $\DS_0$ and the bias in cosmological parameters (\SO)
    due to uncertainties in the \rcc\ modeling is around $1-2\%$ depending on the realism of the mock galaxy catalogs.
	Using the mock galaxy catalogs with a different non-linear physics (higher \rcc) compared to the mock catalogs
    from which the priors
	on \rcc\ are derived, the bias in \SO\ can be $\sim2\%$. 
    These biases are reduced if we choose a larger $r_0$, at the cost of losing some small-scale information.
	We also test the possible impact of baryonic physics on our models using a toy model based on Illustris
	simulations \citep{Genel2014,Dai2018} to modify the small scale
	galaxy-matter cross correlations.  Our results suggest that most of the impact of baryonic physics is absorbed in the nuisance parameters, but
    the impact on $S_8$ can still be up to $2\%$.

	We also derived the constraints on cosmological parameters using data from BOSS spectroscopic samples and galaxy lensing maps from SDSS and CMB
	lensing maps from Planck.
    In comparison with predictions using the Planck cosmology ($\sigma_8=0.8228$, $\Omega_m=0.307$),
    we find our measurements to be lower, with the tension on
	\SO\ being of order 20\%, $2-4\sigma$ (stat) for the case of galaxy lensing while CMB lensing being lower by $10\%$ ($\sim 1\sigma$). Our
    results are
	consistent with recent lensing measurements from the DES and KiDS surveys at the $\sim2\sigma$ level or better.
	\cite{Leauthaud2017} also observed the small scale ($r_p<10\mpch$) galaxy-lensing cross correlations using
	CFHTLenS and BOSS-CMASS
	galaxies to be lower than the predictions from the galaxy mock catalogs designed to match the clustering of
	CMASS galaxies. They performed test for theoretical uncertainties in modeling the measurements, including
	the effects of baryons, assembly bias as well as extensions to the base \lcdm\ model and found that while
	any single effect cannot explain the size of the discrepancy, these effects when combined together can
	lead to differences of the order of observed discrepancy.
	We also estimate systematic
	uncertainties for our measurements, with up to $\sim 2-3\%$ uncertainty from \referee{small scale modeling}, $\sim 2\%$
	shear calibration uncertainty (multiplicative bias)
	and $\lesssim5\%$ due to photo-z calibration uncertainties \citep{Singh2018eg}. Combined, the systematic
	error budget on galaxy lensing measurements
	is $\sim6\%$, which is comparable to the statistical errors on the measurements. There can also be further
	systematic contamination in galaxy lensing measurements due to intrinsic
	alignments, though these have been shown to be consistent with zero at the $5\%$ level \citep{Blazek2012} for the
	galaxy lensing sample used in this work.

    It is difficult to discern the implications of the tension between our measurements
	 and Planck. It is possible that this is a statistical
     fluctuation, and it is also possible that the actual amplitude is a bit lower
     than the central Planck value. Another possibility is that the \lcdm\ model is incomplete and 
     extensions including the effects of neutrinos, modified gravity models or extended dark energy 
     models can explain some these tensions. 
	 More concerning is the impact of systematic
     and modeling errors on this method,
     which remain somewhat difficult to quantify, and which are large comparable to
     the statistical error. To some extent this is by design:
     there is a trade-off between the
     statistical and modeling errors.  As we push to smaller scales statistical errors
     decrease and modeling errors increase, and one chooses the scale where the two
     are comparable. Still, the optimal
     choice of this scale remains somewhat model dependent.
     Even though we have removed all the
     information from scales below 1-2~Mpc$/h$ and we have quantified the
     modeling error by comparing to the impact of HOD models on the
     analysis, it is possible that with a broader HOD
     parametrization we would need to revisit this.
     While future measurements with ongoing surveys such as DES, KiDS and upcoming surveys like LSST and Euclid
	 will help in better understanding of systematics,
     our results highlight the importance of further improving
	 our understanding of theoretical modeling at small scales.

\section*{Acknowledgments}
We thank Martin White and Simone Ferraro for useful discussions. We also thank Shadab Alam for providing chains for BOSS DR12 analysis and
Chiaki Hikage for providing HSC chains. 

This work was supported by NASA grant NNX15AL17G.
RM is supported by the Department of Energy Cosmic Frontier program, grant DE-SC0010118.

Funding for SDSS-III has been provided by the Alfred P. Sloan Foundation, the Participating Institutions, the National Science Foundation, and the U.S. Department of Energy Office of Science. The SDSS-III web site is http://www.sdss3.org/.

SDSS-III is managed by the Astrophysical Research Consortium for the Participating Institutions of the SDSS-III Collaboration including the University of Arizona, the Brazilian Participation Group, Brookhaven National Laboratory, Carnegie Mellon University, University of Florida, the French Participation Group, the German Participation Group, Harvard University, the Instituto de Astrofisica de Canarias, the Michigan State/Notre Dame/JINA Participation Group, Johns Hopkins University, Lawrence Berkeley National Laboratory, Max Planck Institute for Astrophysics, Max Planck Institute for Extraterrestrial Physics, New Mexico State University, New York University, Ohio State University, Pennsylvania State University, University of Portsmouth, Princeton University, the Spanish Participation Group, University of Tokyo, University of Utah, Vanderbilt University, University of Virginia, University of Washington, and Yale University.

\bibliographystyle{mnras}
  \bibliography{sukhdeep}

\appendix
	\section{Analysis with different sets of mock galaxy catalogs }
	\label{appendix:mocks}
	In this appendix we show the result of applying our analysis method to a different set of mock galaxy catalogs
    that were generated by \cite{Reid2014} (Med-Res
	mocks) to match the
	clustering of the BOSS galaxy samples.

\begin{figure*}
			\begin{subfigure}{\columnwidth}
        		\centering
         		\includegraphics[width=\columnwidth]{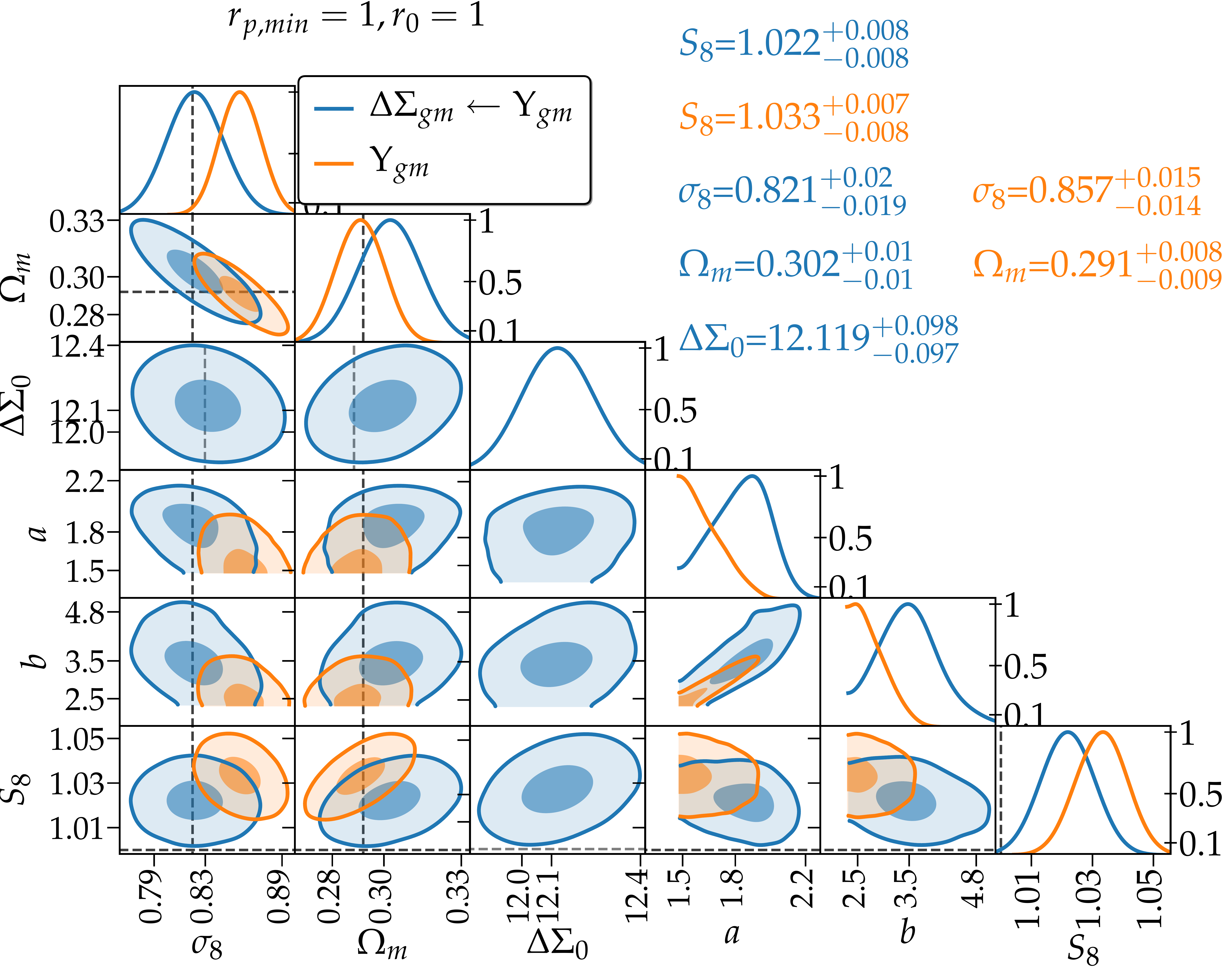}
				\caption{}
	    	     \label{fig:Beth_sim_r01}
		     \end{subfigure}\hfill
		     \begin{subfigure}{\columnwidth}
        		\centering
		         \includegraphics[width=\columnwidth]{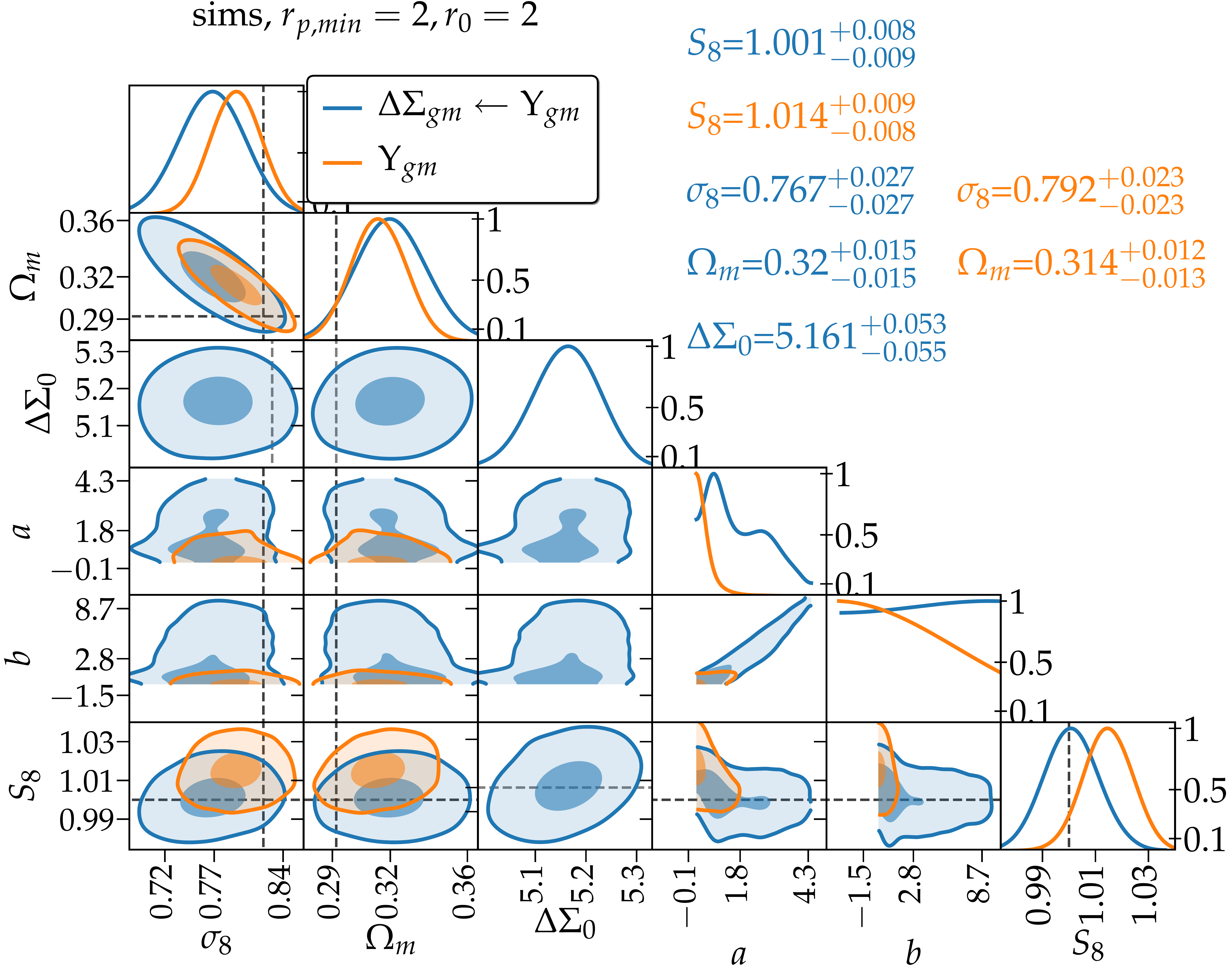}
				\caption{}
	    	     \label{fig:Beth_sim_r02}
		     \end{subfigure}\hfill
		     \caption{Parameter constraints for a different set of mock catalogs, Mid-Res in \citealt{Reid2014},
             using the same \rcc\ priors as for the
		     	main analysis. For $r_0=1 \mpch$ (left panel), the \SO\ constraints are biased by $\sim2-3\%$
                ($\sim3\sigma$), but for $r_0=2\mpch$ (right panel) we recover
				the correct cosmological parameters. Note that for this plot, we defined
				$\SO=\left(\frac{\sigma_8}{0.82}\right)^{0.8}\left(\frac{\Omega_m}{0.292}\right)^{0.6}$,
				where the $\sigma_8$ and $\Omega_m$ normalization
				factors correspond to the cosmology of simulations used to generate these mock catalogs.
		      }
		     \label{fig:Beth_sim}
    \end{figure*}

	In the analysis use the same \rcc\ priors that were derived from the mock galaxy catalogs used in the main part
    of the paper. The mock catalogs from \cite{Reid2014} have
	slightly different (higher) non-linear effects compared to the main mock galaxy catalogs, as was shown in
    \cite{Singh2018eg}. As a result, in Fig.~\ref{fig:Beth_sim} when using a
	low $r_0$ value of $1\mpch$, the results are biased by $\sim 2\%$, while the bias is negligible for
    $r_0=2\mpch$.
 	For our main results we will quote the cosmological parameters using both $r_0=1\mpch$ and $r_0=2\mpch$.

\begin{figure}
        		\centering
         \includegraphics[width=\columnwidth]{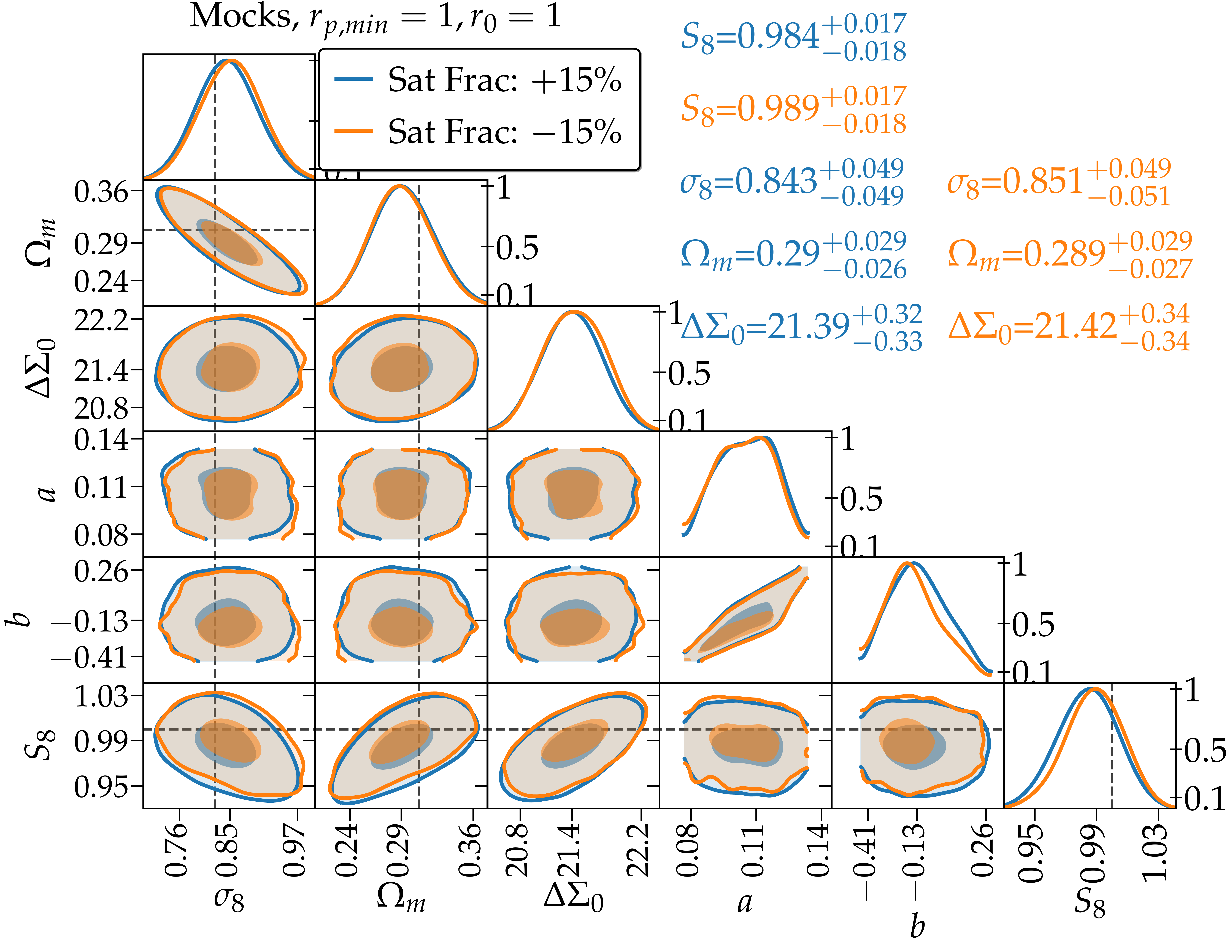}
		  \caption{ Comparison of constraints obtained using the redshift-dependent mock galaxy catalogs described in Section~\ref{ssec:z_mocks}
		  but with different satellite fractions. We obtain
		  		consistent results, within $1\sigma$ of the fiducial values.
		      }
	    \label{fig:evol15}
    \end{figure}

    In Fig.~\ref{fig:evol15} we show the constraints obtained by varying the satellite fraction using the mock
    catalogs described in
	Section~\ref{ssec:z_mocks}. We obtain consistent values of cosmological parameters (within $1\sigma$) with these
    mock catalogs,
    as expected since the primary effect of varying satellite
    fraction will be to change \rcc\ which for these mock catalogs is accounted for in the width of our priors.

    Note choosing higher $r_0$ decreases the bias on \SO\ which is the main parameter constrained using
    galaxy-lensing cross correlations. Due to different scale dependence of \rcc, marginalized constraints on
    $\sigma_8$ and $\Omega_m$ can still be somewhat biased as the wrong scale dependence of \rcc\ is compensated
    by varying $\Omega_m$ values. We observe these biases to be present at low levels ($\lesssim1\sigma$ level on
    $\sigma_8$ and $\Omega_M$), though there could be some contribution from (correlated) noise also in these.
    These biases are much smaller than the statistical uncertainties in our results using lensing maps and
    removal of these biases will require more realistic and precise mock galaxy catalogs than have been used in this
    work.

\section{Impact of covariance}
\label{appendix:covariance}
	    	\begin{figure}
        		\centering
         		\includegraphics[width=\columnwidth]{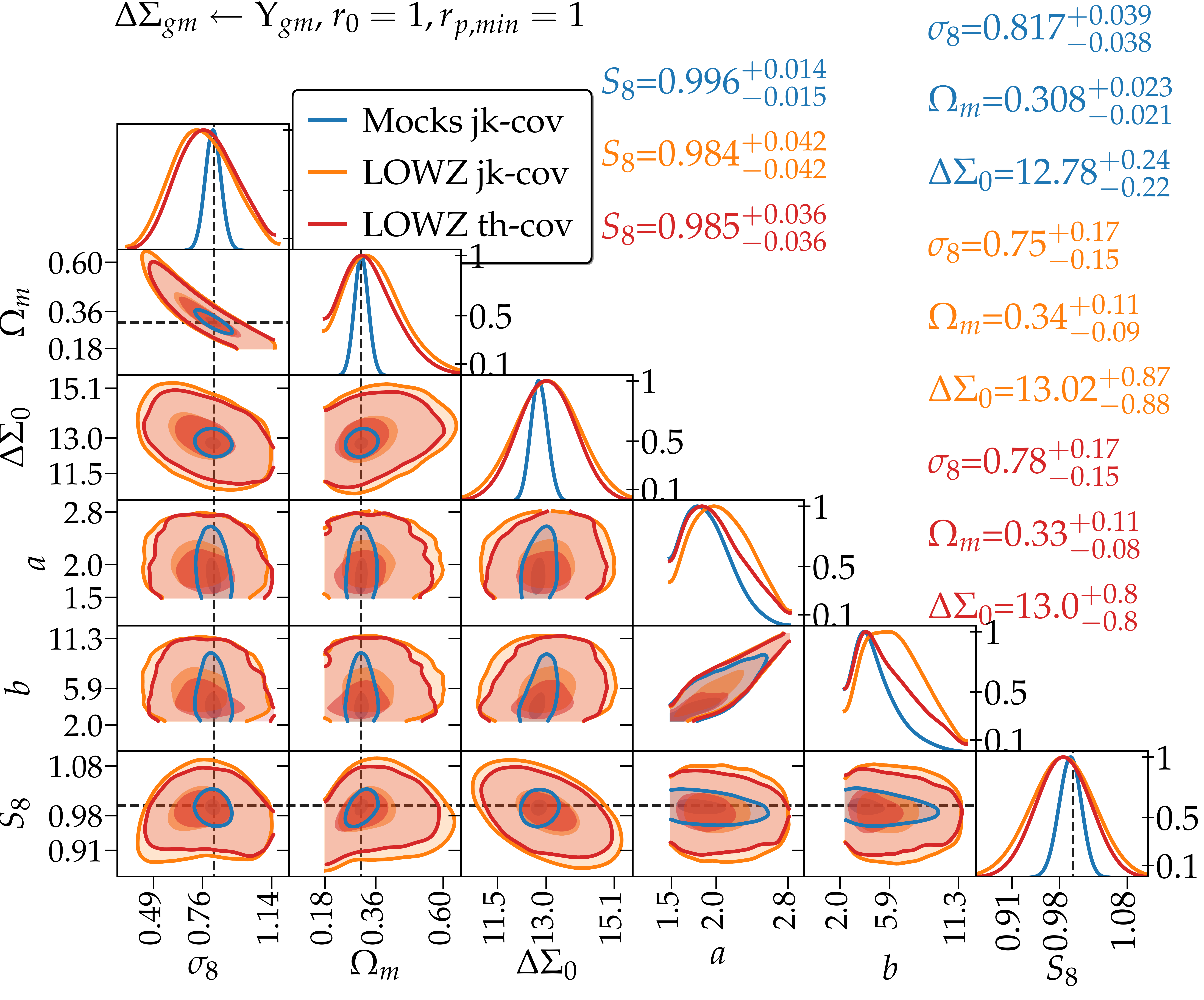}
		         \caption{Comparison of constraints obtained using different covariance matrices:
		         the jackknife covariance from mock catalogs (mocks-jk, no shape noise), the
		         jackknife covariance from the data (LOWZ-jk, includes shape noise) and the theoretical prediction
                 for the data covariance
		         (LOWZ-th). The mock catalogs covariance gives
		         the smallest covariance on parameters since it does not include the shape noise. The data
                 covariance including shape noise increases
		         the parameter
		         errors
                 by a factor of $\sim 3$ as shape noise reduces information especially from small scales.
		         	}
	    	     \label{fig:DS_cov_comp}
    	    \end{figure}
	    	In Fig.~\ref{fig:DS_cov_comp} we show the comparison of constraints obtained using different covariance
            matrices. When using the mock catalogs
			covariance, we get the tightest constraints on the cosmological parameters as the mock catalogs
            covariance does not include shape noise and it also optimally
			weighs the data points. Shape noise increases the
			measurement covariance, especially the diagonal part of the covariance at small scales and hence reduces
            the information from these scales,
			increasing the cosmological parameter covariance. We also compare the impact of using data covariance
            from jackknife and the data covariance from
			theoretical predictions. Since the jackknife covariance is estimated from 100 regions, we apply the
            Hartlap correction \citep{Hartlap2007} to it, which
			increases the parameter errors by $\sim 10\%$.
			There is a further difference between the jackknife and theoretical covariance with the diagonal entries of
			theoretical covariance being lower by $\sim 5\%$.
			As discussed in \cite{Singh2017cov}, this is likely due to some contribution from
			noise in the jackknife covariance and also  the fact that in computing
			the theoretical covariance we do not include the potential contributions from the systematics, which can
            contribute to the jackknife covariance even if
			removed from the mean signal. There can also potentially be some impact from non-Gaussian and super-sample covariance terms.
			Since the mean values of cosmological parameters obtained using different covariances is consistent,
            our main results using the data
            will be computed using the jackknife covariance though we will check the results using theoretical covariance as well.

	    \section{Impact of Baryonic Physics}
        \label{appendix:baryons}
	    	\begin{figure*}
				\begin{subfigure}{\columnwidth}
        			\centering
	         		\includegraphics[width=\columnwidth]{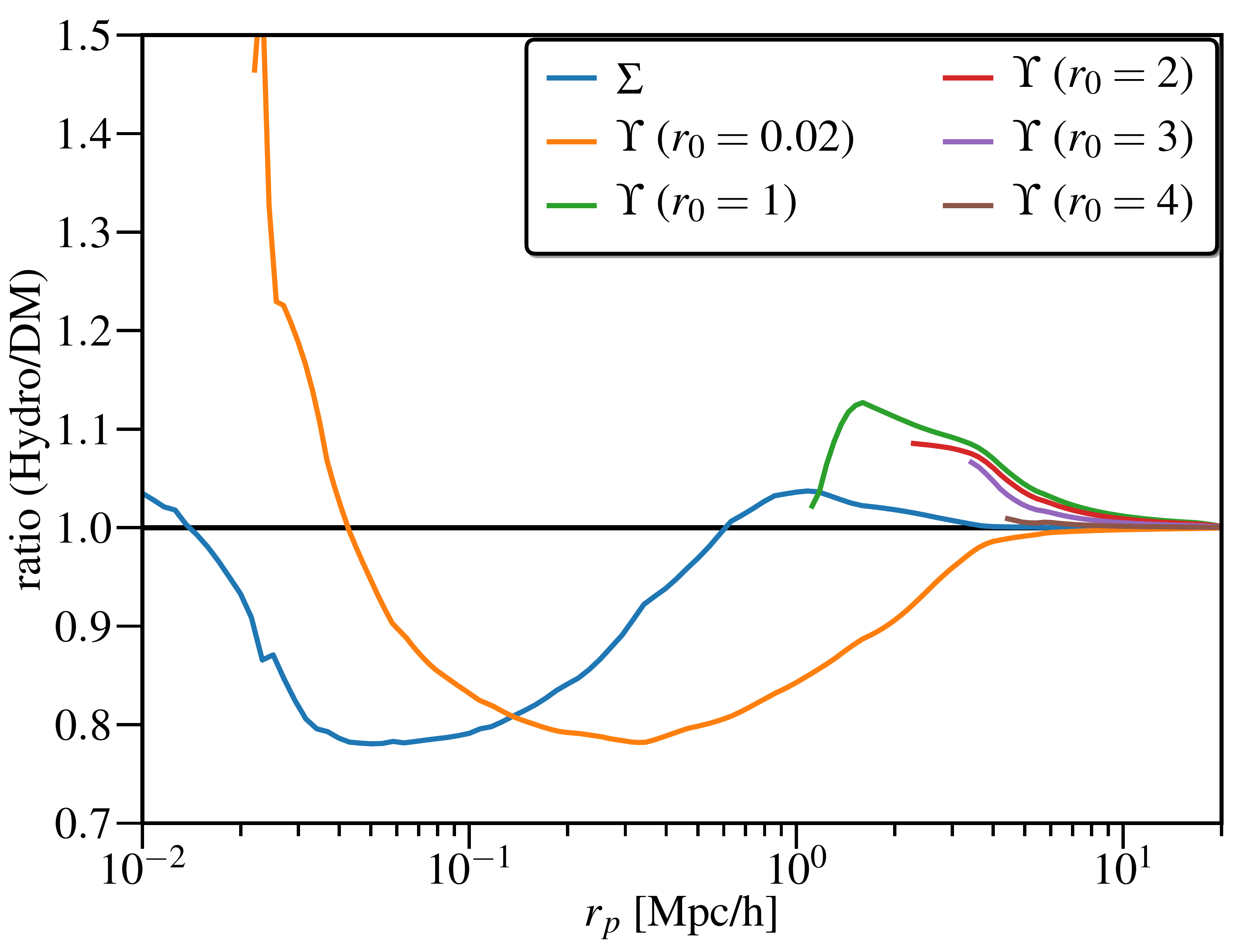}
			         \caption{}
	    		     \label{fig:mocks_baryons_ratio}
			     \end{subfigure}\hfill
			     \begin{subfigure}{\columnwidth}
			    		\centering
	         		\includegraphics[width=1.05\columnwidth]{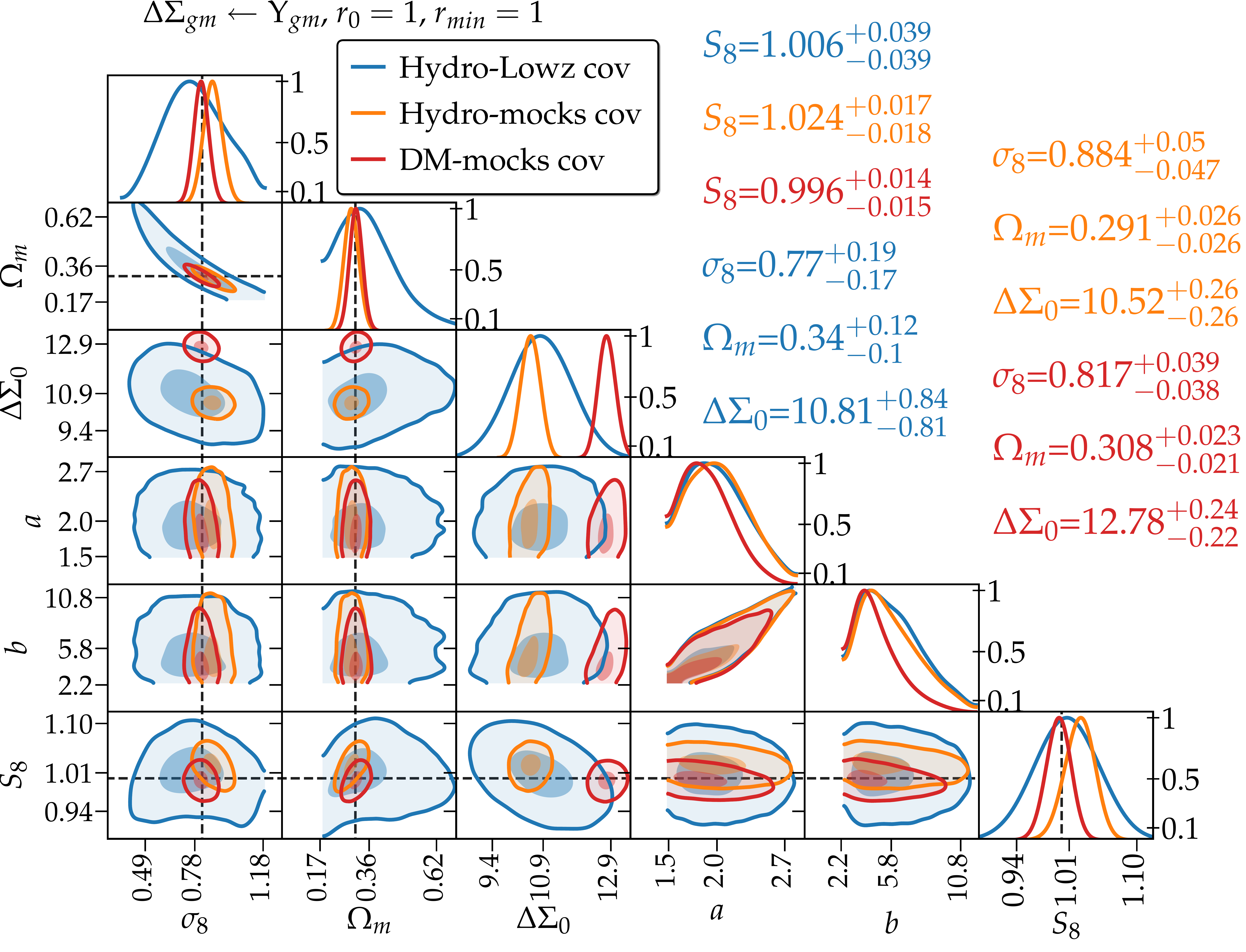}
			         \caption{}
	    		     \label{fig:mocks_baryons_cosmo}
			    \end{subfigure}
			    \caption{a) The ratio of the projected matter distribution (projected galaxy-matter cross correlation)
					in and around halos between Illustris hydrodynamical and dark matter only
                    simulations. We use this ratio to modify
			    	the galaxy-matter cross correlation signal to study the impact of baryonic physics on cosmological parameter inferences. b) Parameter
					inferences when using the galaxy-matter cross correlations from mock catalogs (DM) and the modified galaxy-matter cross correlations using
                    the
					ratio from (a). While the impact of baryonic physics changes the galaxy-matter cross correlations significantly at small scales, most of
                    the
					effects of these changes are absorbed in $\DS_0$ and \rcc.
					When using the mock catalogs covariances, the \SO\ changes by $\sim 2\sigma$ ($\sim 2.5\%$), which suggests that baryonic physics can
					be important. However, when using the data covariance, which downweights the small scales due to shape noise, these effects are much
					smaller.
			}
	    		\label{fig:mocks_baryons}
    	    \end{figure*}

	    	The effects of baryonic physics tend to modify the matter distribution within and around dark matter halos, modifying the small-scale correlations.
			To estimate the impact of such changes, we
			compute the ratio of the projected matter distribution around halos in dark matter-only and Illustris simulations \citep{Genel2014} using halo
			profiles provided by \cite{Dai2018}, as shown in
			Fig.~\ref{fig:mocks_baryons_ratio}. Using
			this ratio, we then modify the galaxy-matter cross correlations computed from the mock catalogs  and then perform the cosmological analysis using the
			mock catalog
			covariance. As shown in Fig.~\ref{fig:mocks_baryons_cosmo}, most of the impact of baryons is absorbed in $\DS_0$ and \rcc, and the
			constraints on \SO\ change by $\sim 2.5\%$.
			When using the data covariance, the shape noise
			downweights the small scales more and the impact of baryonic physics is further reduced in the cosmological parameter. We also note that we are using
			the
			Illustris simulation, for which the AGN feedback is likely
            too strong \citep{Springel2018},
            and hence has a significant impact of baryons out to scales
            larger than seen in other hydrodynamical simulations. We expect the impact of
            baryonic physics to be smaller than this estimate. Nevertheless, the results
            in
			Fig.~\ref{fig:mocks_baryons}
			suggest that future works using better weak lensing measurements may need to model the baryonic physics more carefully in order to achieve
			unbiased constraints at better than $1\%$ accuracy.

	\section{Effects of binning}
		\label{appendix:binning_effect}
\begin{figure}
        		\centering
         		\includegraphics[width=\columnwidth]{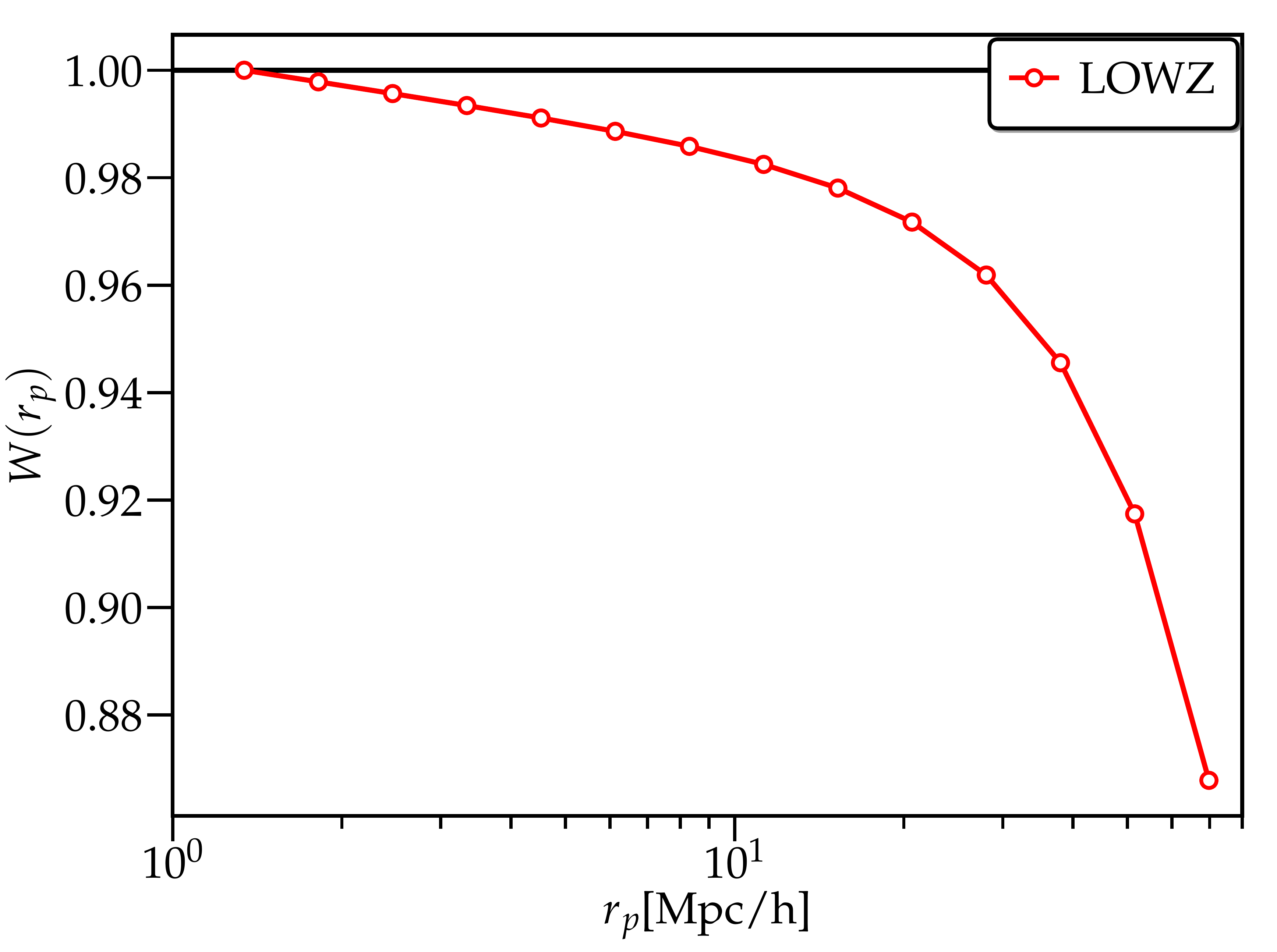}
		     \caption{The survey window function $W(r_p)$ for \DS\ measured using random-source pair counts, with the value at the smallest bin normalized to
             1.
		      }
	    	     \label{fig:DS_window}
    	\end{figure}

		The measured correlation function in bins is an average weighted by the available number of pairs at each scale.
		The number of pairs scale with the
		volume in the bin modulated by the effects of the survey window function. The binned projected correlation function is
		\begin{align}
		\widehat{\DS}_i=\frac{\int^{r_{p,high}}_{{r_{p,low}}}\mathrm{d}r_p'r_p' W(r_p')\DS(r_p')}{\int^{r_{p,high}}_{{r_{p,low}}}\mathrm{d}r_p'r_p' W(r_p')}
		\end{align}
		where $\DS_i$ refers to the \DS\ value measured in bin $i$ with $r_{p,high},r_{p,low}$ marking the limits of the bin. $W(r_p)\in[0,1]$ is
		the effect of the survey window, with $W(r_p)<1$ implying that some pairs at $r_p$ are missed relative to the expected scaling
		due to the effects of the window. We estimate
		$W(r_p)$ using the pair counts involving random lenses (random-source pairs for galaxy-lensing cross correlations) in the bins as
		\begin{align}
			W(r_p)=\frac{\widehat{N}_p}{N_\text{lens}\overline{n}_\text{source}\int^{r_{p,high}}_{{r_{p,low}}}\mathrm{d}r_p'r_p'}
		\end{align}
		where $\widehat{N}_p$ is the measured number of pairs and denominator is the expected number of pairs ignoring the window effects.
		When estimating $W(r_p)$, we divide the number of pairs with the bin volume $\int dr_p
		r_p$ and then set the $W(r_p)$ at the smallest scale to 1. This does not affect our calculations as the absolute normalization of $W(r_p)$
		within the bin cancels out when binning the theory (this is not true for covariance) and the effects we study here are due to the gradient of $W(r_p)$ within the bin.
		Finally we fit a smooth function to the estimated $W(r_p)$ in the bins to
		get $W(r_p)$ as function of scale, $r_p$. $W(r_p)$ for galaxy-galaxy
		lensing using the LOWZ sample is shown in Fig.~\ref{fig:DS_window}.

\begin{figure}
        		\centering
         		\includegraphics[width=\columnwidth]{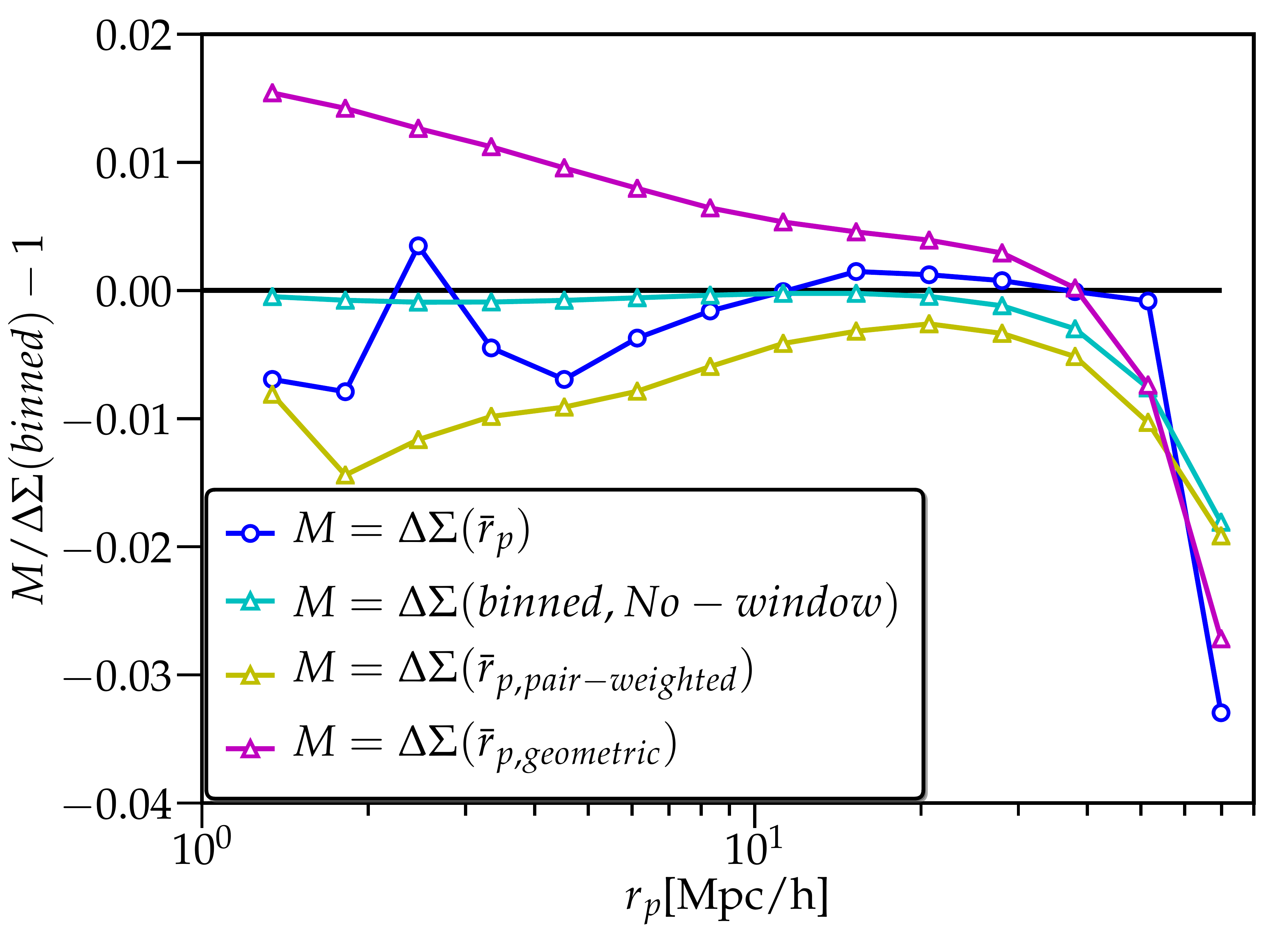}
		     \caption{Comparison of the binned theoretical model with the binned model without accounting for effects of window (cyan) and
		     the model computed at different definitions of the central $r_p$ of the bin as described in the Eq.~\ref{eq:rp_mean}-\ref{eq:rp_pw}.
		      }
	    	     \label{fig:DS_window_ratio}
    	\end{figure}

		In Figure~\ref{fig:DS_window_ratio} we show the differences between the binned theoretical predictions and those computed at various
		effective scales in the bins, including the arithmetic mean  ($\overline{r}_p$), the geometric mean
		($\overline{r}_{p,\text{geometric}}$),

  		and the pair-weighted mean
		($\overline{r}_{p,\text{pair-weighted}}$) defined as
		\begin{align}
		&\overline{r}_p=\frac{r_{p,high}+r_{p,low}}{2}\label{eq:rp_mean}\\
		&\overline{r}_{p,\text{geometric}}=\sqrt{r_{p,high}\times r_{p,low}}\\
		&\overline{r}_{p,\text{pair-weighted}}=\frac{\int^{r_{p,high}}_{{r_{p,low}}}\mathrm{d}r_p'r_p'^2 W(r_p')}{\int^{r_{p,high}}_{{r_{p,low}}}\mathrm{d}r_p'r_p'W(r_p')}\label{eq:rp_pw}
		\end{align}
		The impact of different choices of effective $r_p$ can be understood by the fact that the effective scale of the measurement
		within the bin depends on the scaling of the pair counts (which also depends on survey window) as well as the scaling of the correlation
		function. For example, if the function scales as $r_p^{-1}$ and ignoring the effects of the survey window the pair counts scale as $r_p$, it can be shown that the
		measurement
		is effectively made at the arithmetic mean $\overline{r}_p$.
		The scaling of \DS, $\partial\log\DS/\partial \log r_p\sim-1$ (but varies
		with $r_p$), explains the result that the theoretical prediction evaluated at the arithmetic mean $\overline{r}_p$ is close to the binned theory, though there
		are some variations due to the varying slope of \DS. The pair-weighted $r_p$ overestimates the effective scale and thus under-predicts
		the theory while the geometric mean has the opposite effect.
		Since the survey window tends to down-weight the larger $r_p$,
		the effective scale will be smaller than $\overline{r}_p$ for the bins in which the survey window has a large gradient. Thus if we evaluate the model at $\overline{r}_p$, we will be under-predicting
		the value of the correlation function within the bin. This explains the large discrepancy at larger $r_p$ between the binned models with and
		without the window effects as well as the model computed at $\overline{r}_p$.

		We also note that these effects depend on the size of the bins, with less differences between the choices of effective scale when using smaller bins.
		While smaller bins may not always be desirable (e.g., covariance estimation is more challenging if there are too many bins), one can always perform
		the measurement in small bins and then rebin the measurements with the desired
 		weights (which will also require proper treatment in
		covariance). Also, the choice of effective scale depends on the scaling of the function being measured, and
		there is no unique choice of effective $r_p$ that works for all functions. The best (and the correct) approach is to bin the theory
		predictions with proper weights.

\begin{figure*}
			\begin{subfigure}{\columnwidth}
        		\centering
         		\includegraphics[width=\columnwidth]{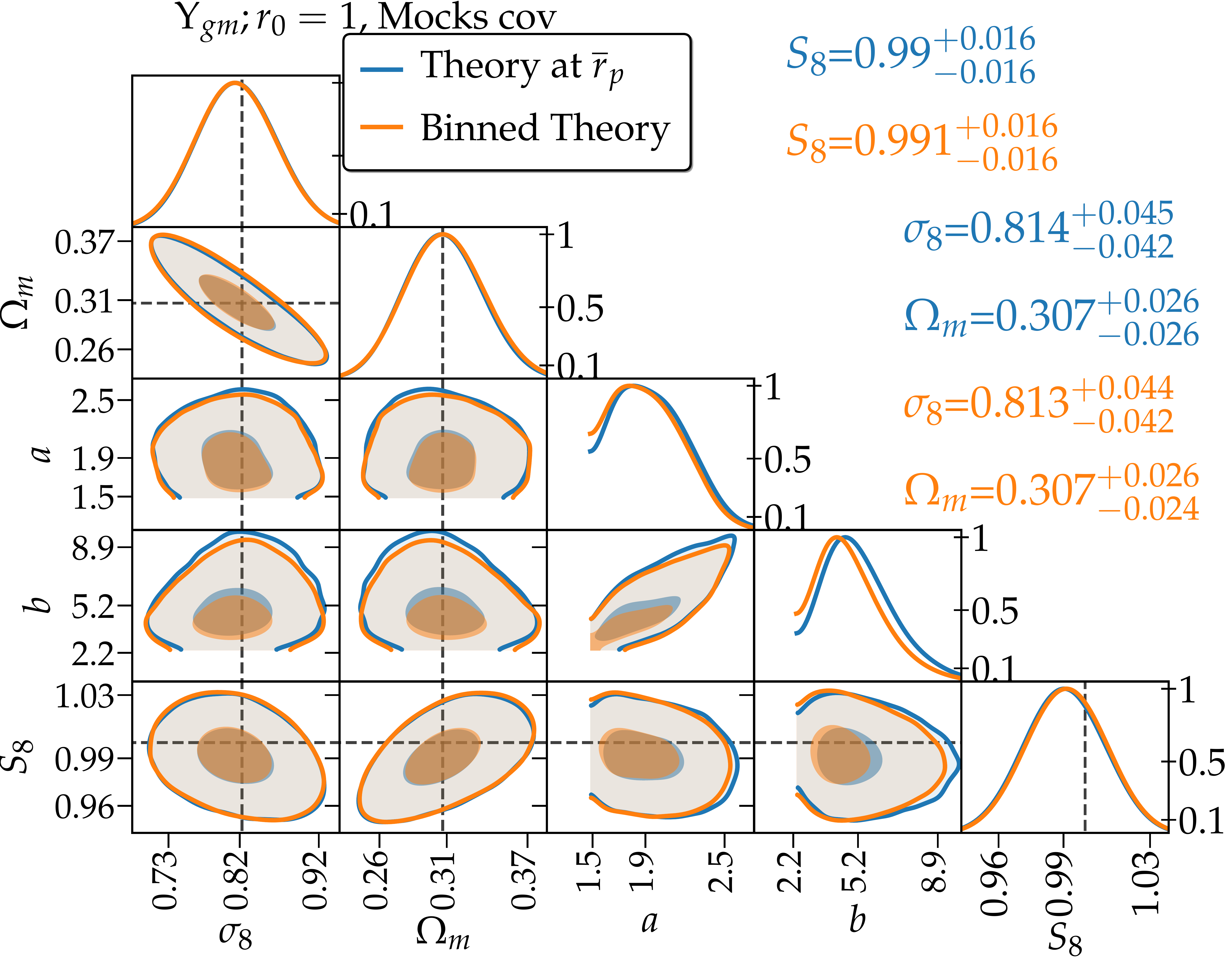}
				\caption{}
	    	     \label{fig:Upsilon_theory_bin}
		     \end{subfigure}\hfill
		     \begin{subfigure}{\columnwidth}
        		\centering
		         \includegraphics[width=\columnwidth]{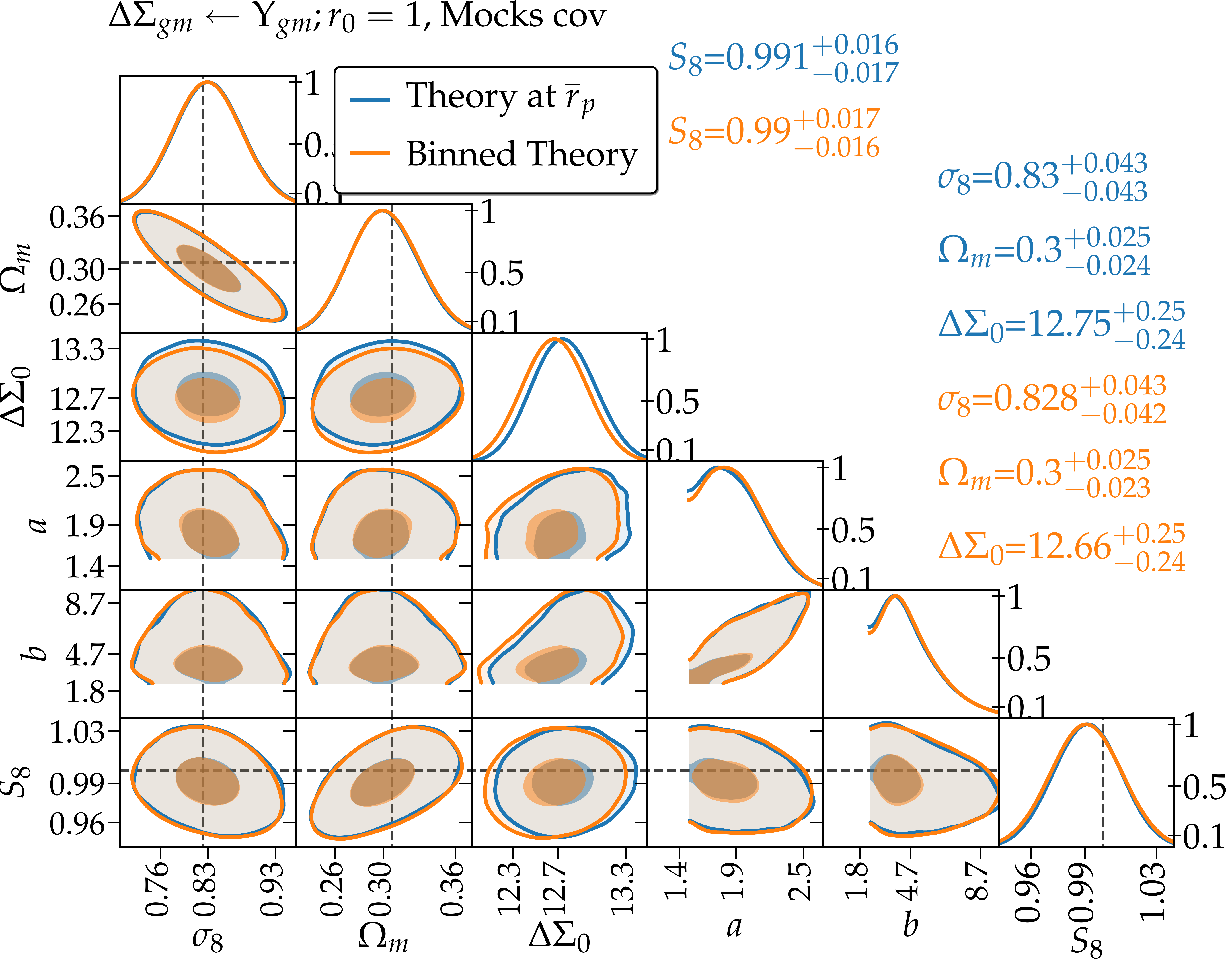}
				\caption{}
	    	     \label{fig:DS_theory_bin}
		     \end{subfigure}\hfill
		     \caption{Comparison of cosmological parameter constraints obtained by binning the theory and computing theory at $\overline{r}_p$. We find negligible
			     difference in the \SO\ constraints in the two cases, with some differences getting absorbed in the nuisance parameters.
		      }
		     \label{fig:theory_bin}
    	    \end{figure*}

		Finally, in Figure~\ref{fig:theory_bin},
		we show the impact of computing the theoretical prediction at $\overline{r}_p$ compared to the binned theory with mask effects. We find that the final impact
		on \SO\ is $\sim0.1\%$ (or $<0.1\sigma$) on our results with some of the effects on small scales getting absorbed in the \rcc\ and $\DS_0$
		parameters.

\end{document}